\documentclass{article}
\usepackage{arxiv}
\usepackage{multirow} 
\usepackage{tabularx}
\usepackage{makecell}
\usepackage[utf8]{inputenc} 
\usepackage[T1]{fontenc}    
\usepackage{hyperref}       
\usepackage{url}            
\usepackage{booktabs}       
\usepackage{amsfonts}       
\usepackage{nicefrac}       
\usepackage{microtype}      
\usepackage{graphicx}
\usepackage{natbib}
\usepackage{doi}

\title{The Effect of Limited Mobility on the Experienced Segregation of Foreign-born Minorities}

\author{ \href{https://orcid.org/0000-0002-6982-1654}{\includegraphics[scale=0.06]{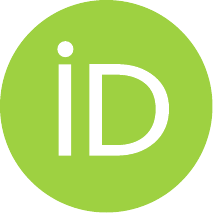}\hspace{1mm}Yuan~Liao}\thanks{Corresponding author. Also affiliated with Department of Applied Mathematics and Computer Science, Technical University of Denmark, Lyngby, Denmark.} \\
	Department of Space, Earth and Environment\\
	Chalmers University of Technology\\
	Gothenburg, Sweden \\
	\texttt{yuan.liao@chalmers.se} \\
	\And
	\href{https://orcid.org/0000-0001-6671-2578}{\includegraphics[scale=0.06]{orcid.pdf}\hspace{1mm}Jorge~Gil} \\
	Department of Architecture and Civil Engineering\\
	Chalmers University of Technology\\
	Gothenburg, Sweden\\
	\texttt{jorge.gil@chalmers.se} \\
    \And
	\href{https://orcid.org/0000-0002-4852-1177}{\includegraphics[scale=0.06]{orcid.pdf}\hspace{1mm}Sonia~Yeh} \\
	Department of Space, Earth and Environment\\
	Chalmers University of Technology\\
	Gothenburg, Sweden\\
	\texttt{sonia.yeh@chalmers.se} \\
    \And
	\href{https://orcid.org/0000-0003-2125-7465}{\includegraphics[scale=0.06]{orcid.pdf}\hspace{1mm}Rafael~H. M.~Pereira} \\
	Institute for Applied Economic Research (Ipea) - Brazil\\
	Data Science Division, Brazil\\
	\texttt{rafael.pereira@ipea.gov.br} \\
    \And
	\href{https://orcid.org/0000-0001-6003-1165}{\includegraphics[scale=0.06]{orcid.pdf}\hspace{1mm}Laura~Alessandretti} \\
	Department of Applied Mathematics and Computer Science\\
	Technical University of Denmark\\
    Lyngby, Denmark\\
	\texttt{lauale@dtu.dk} \\ 
}

\renewcommand{\shorttitle}{The Effect of Limited Mobility on the Experienced Segregation of Foreign-born Minorities}

\hypersetup{
pdftitle={A template for the arxiv style},
pdfsubject={q-bio.NC, q-bio.QM},
pdfauthor={David S.~Hippocampus, Elias D.~Striatum},
pdfkeywords={First keyword, Second keyword, More},
}

\begin{document}
\newcommand{\tabincell}[2]{
\begin{tabular}{@{}#1@{}}#2\end{tabular}
}
\maketitle

\begin{abstract}
Segregation is a key challenge in promoting more diverse and inclusive cities. 
Research based on large-scale mobility data indicates that segregation between majority and minority groups persists in daily activities beyond residential areas, like visiting shops and restaurants. 
Aspects including lifestyle differences, homophily, and mobility constraints have been proposed as drivers of this phenomenon, but their contributions remain poorly quantified.  
Here, we elucidate how different mechanisms influence segregation outside home, looking at the distinctive segregation experienced by native and foreign-born individuals. 
Our study is based on the movement of $\sim$320,000 individual smartphone devices collected in Sweden, where immigration creates profound divides. 
We find that while day-to-day activities lead to mixing for native-born individuals, foreign-born individuals remain segregated in their out-of-home activities. 
Using counterfactual simulations, we show that this heterogeneous effect of mobility on experienced segregation results mainly from two mechanisms: homophily and limited travel, i.e. foreign-born individuals (i) tend to visit destinations visited by similar individuals, and (ii) have limited mobility ranges. 
We show that homophily, as represented by destination preference, plays a minor role, while limited mobility, associated with reduced transport access, limits opportunities for foreign-born minorities to diversify their encounters. 
Our findings suggest that enhancing transport accessibility in foreign-born concentrated areas could reduce social segregation.
\end{abstract}


\section{Introduction}
Segregation based on country of origin is a critical problem in many cities, which has been shown to perpetuate social and economic inequalities \citep{musterd2017socioeconomic, chetty2022social,sousa2022quantifying}. 
The issue is especially acute in countries that have experienced substantial influxes of refugees and migrants in a short period, and has triggered public debates and political attention worldwide \citep{iceland2010residential, malmberg2013segregation,jarvis2023assortative,algan2012cultural}. \par

Research in urban segregation has intensified and expanded in the last decade \citep{liao2024socio}.
Traditionally, studies have focused on \emph{residential segregation} \citep{duncan1955methodological, feitosa2007global, barros2018uneven}, exploring how people from diverse backgrounds spatially distribute in their residence. 
Recent research enabled by detailed smartphone data has shifted towards examining experienced segregation, which also considers people's exposure to different groups in other spaces, such as workplaces, shops, or leisure areas \citep{kwan2013beyond,moro2021mobility,li2022towards}. 
Understanding \emph{experienced segregation} beyond residential boundaries is critical for developing more effective and nuanced interventions to reduce socio-spatial segregation, compared to altering residential location patterns. \par

Empirical work leveraging smartphone data has revealed that segregation persists in people's day-to-day activities \citep{moro2021mobility,nilforoshan2023human,xu2024experienced}. 
The reasons underlying the segregation outside home, however, have remained elusive. 
Researchers have hypothesized a few key possible mechanisms. 
One of such reasons would be \emph{lifestyle}, with individuals from different backgrounds visiting different types of venues (e.g., restaurants, shops, museums) during their daily activities \citep{moro2021mobility,yang2023identifying}. 
A second reason would be \emph{homophily}, according to which individuals gravitate towards people with similar backgrounds \citep{xu2022beyond,huang2022unfolding}. 
Homophily would manifest itself as destination preference \citep{chen2024behavioral}, driving individuals to visit certain places over others. 
Another reason would be \emph{limited mobility ranges}, with individuals mainly participating in daily activities close to home. 
This could result either from limited transportation access \citep{vachuska2023racial} or sufficient local services \citep{abbiasov202415}.
Finally, the urban structure itself, e.g., the geographical distribution of residences and amenities, may shape the places people visit \citep{netto2015segregated,liao2024socio}.
Minorities often reside in disadvantaged neighborhoods, limiting opportunities for co-presence with other groups \citep{tao2020influence} and the slow evolution of the built environment entrenches urban communities in persistent settlement patterns \citep{patias2023local}.
The effect of the above mechanisms on the segregation experience of different groups remains unclear, partly due to the difficulties separating these intertwined factors. \par

Here, we establish the influence of these mechanisms on experienced segregation using statistical techniques in simulated counterfactual scenarios. 
Our analysis focuses on Sweden, where the issue of segregation by country of birth, i.e., between native-born and foreign-born outside Europe, is particularly critical. 
The percentage of the foreign-born population has increased from 11\% to 21\% between 2000 and 2022, making Sweden the country with the second-largest foreign-born population share among all OECD (Organisation for Economic Co-operation and Development) countries \citep{oecd2022international}.
Our study is based on a large-scale smartphone dataset that captures $\sim30$ million activity geolocations for $322,919$ individual devices over seven months in 2019.

\section{Results}
\subsection{The Uneven Impact of Daily Mobility on Segregation}
To establish how segregation is affected by daily travel for out-of-home activities, we estimate for each individual their (i) \emph{residential segregation} and their (ii) \emph{experienced segregation}. 
Note that we use post-stratification techniques to mitigate the effect of sampling and population biases in our dataset (see more details in the Methods section, Reducing sampling and population biases). \par 

\begin{figure*}[!tbhp]
\centering
\includegraphics[width=1\textwidth]{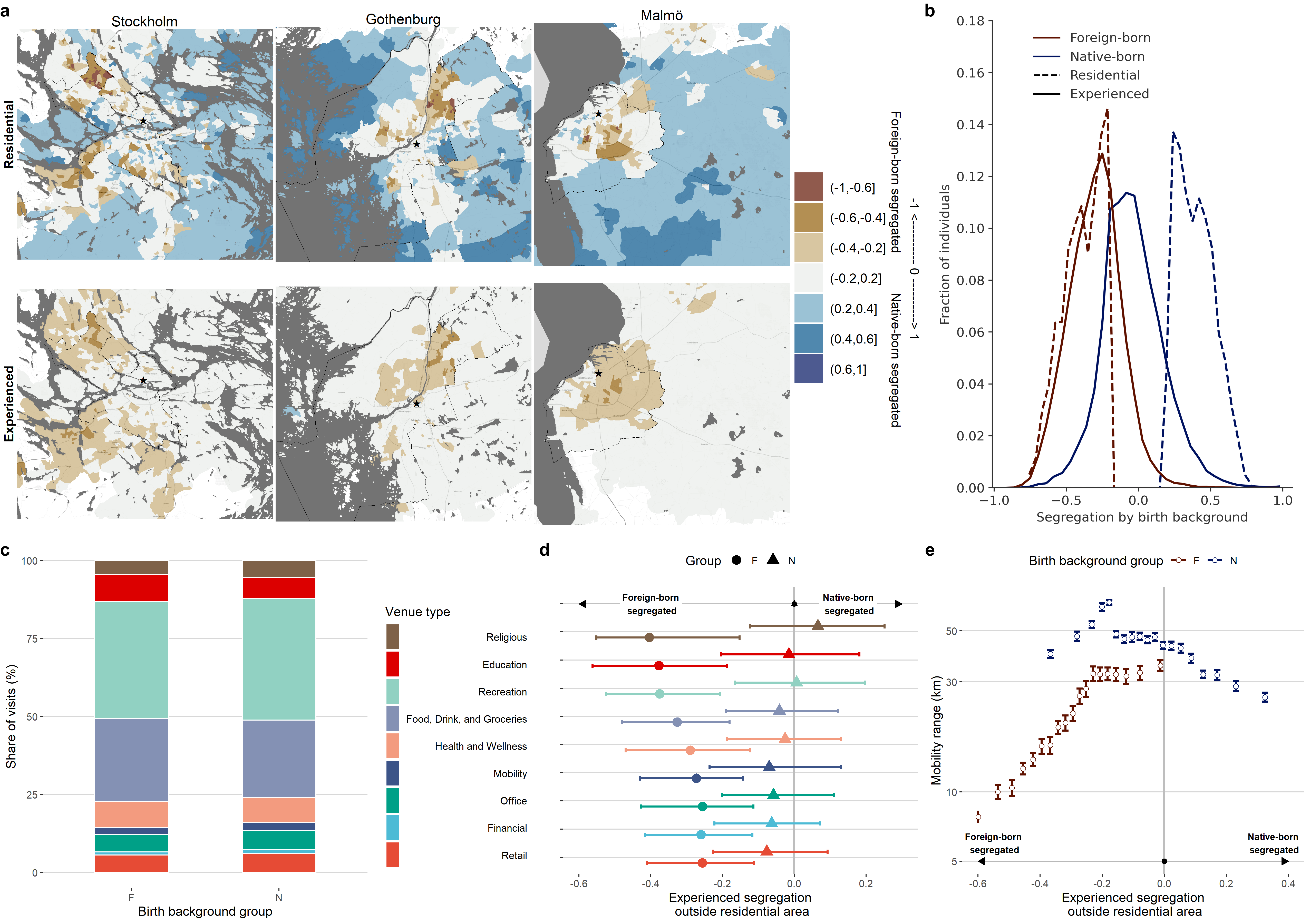}
\caption{\textbf{Residential segregation and experienced segregation outside the residential area}. \textbf{a}, Choropleth maps showing residential (top row, $ICE_r$) and experienced (bottom row, $ICE_e$) segregation in Sweden's three major cities: Stockholm, Gothenburg, and Malmö (left to right). Colors capture the average segregation within each statistical area. Black lines show municipality boundaries. \textbf{b}, The distribution of residential (dashed lines) and experienced (full lines) segregation for individuals in the foreign-born (red) and native-born (blue) segregated groups. Individuals are categorized by residential segregation into three groups: $ICE_r < -0.2$ (foreign-born), $ICE_r > 0.2$ (native-born), and $-0.2 \leq ICE_r \leq 0.2$ (mixed). \textbf{c}, Share of visits by birth background group and venue type. \textbf{d}, Experienced segregation by birth background group and venue. Error bars show the ranges between the 25th and 75th percentiles. \textbf{e}, Mobility range of individuals by their experienced segregation. The mobility range is represented by the radius of gyration (km). Lines show the median errors. Red lines/dots are the results for those living in foreign-born segregated areas (Group F), and blue lines/dots are for native-born segregated areas (Group N), divided into 19 groups based on the quantile breaks of their respective segregation level. }
\label{fig:seg_disp_fig1}
\end{figure*}

We define individuals' \emph{residential segregation}, denoted as $ICE_r$, using the Indicator of Concentration at Extremes (see the Methods section, Segregation indicator) computed within their demographic statistics area of residence (see the Methods section, Stay and Home Detection, and Socioeconomic Attributes Assignment). 
Intuitively, $ICE_r$ reaches its most extreme values for individuals residing solely among foreign-born ($ICE_r=1$) or native-born ($ICE_r=-1$). 
A value of $ICE_r=0$ indicates a residential composition that mirrors the national average in Sweden -- comprising 80.4\% native-born, 11.1\% foreign-born outside Europe, and 8.5\% other (foreign-born in Europe).
We categorise areas with $|ICE_r|>0.2$ as segregated, because they deviate significantly from an expected distribution where residency is independent of birth background, at $p<0.01$ (see the Methods section). 
Consequently, we classify individuals into three groups: those for which $ICE_r<-0.2$, living in native-born segregated areas (N); those with $ICE_r>0.2$, living in foreign-born segregated areas (F); and those with $-0.2\leq ICE_r \leq0.2$ living in mixed areas (M).
We find pronounced residential segregation in Sweden, with 40.8\% of individuals living in areas with a higher percentage of native-born than the national average (Group N), 18.4\% living in areas with a higher percentage of foreign-born (Group F), and 40.8\% in mixed areas (Group M).
Segregation is especially marked in Sweden's three largest municipalities -- Stockholm, Gothenburg, and Malmö -- where over 40\%, 49\%, and 45\% of areas respectively, exhibit a significant average level of residential segregation, $|ICE_r|>0.2$ (see Fig. \ref{fig:seg_disp_fig1}a, top row). \par

For each individual, we assess \emph{experienced segregation}, denoted as $ICE_e$, by measuring the composition of the people encountered during activities outside the home on non-holiday weekdays (for further details, see the Methods section, Residential and experienced segregation). 
The focus on non-holiday weekdays narrows the scope of our analysis to understand the role of routine day-to-day activities.  
Encounters are defined as co-locations within the same 30-minute interval at the same area with an average size of $0.087 km^2$.
Also, in this case, $ICE_e$ is measured as the Indicator of Concentration at Extremes: $ICE_e=0$ indicates that the demographic composition of the encounters aligns with the overall national composition, while extreme values $ICE_e=1$ and $ICE_e=-1$ occur if encounters are exclusively with either foreign-born or native-born individuals, respectively.
Experienced segregation, although correlated with residential segregation (corr. coefficient: $0.48$), is generally lower (see Fig. \ref{fig:seg_disp_fig1}a). 
Specifically, $ICE_e$ is lower than $ICE_r$ for 62\% of individuals in our dataset. 
We find that 28\%, 40\%, and 73\% of areas in Stockholm, Gothenburg, and Malmö respectively, exhibit high experienced segregation levels, $|ICE_e|>0.2$.
Interestingly, in contrast to the patterns observed for residential segregation (see Fig. \ref{fig:seg_disp_fig1}a, top row), $ICE_e$ is always negative in these cities (see Fig. \ref{fig:seg_disp_fig1}a, bottom row). 
This suggests that mobility may affect the native and foreign-born groups differently. \par

Figure \ref{fig:seg_disp_fig1}b illustrates the shift in segregation distribution -- from residential to outside home experiences -- for each group.
We find that for the native-born individuals (Group N), the average segregation becomes significantly lower from $0.385 \pm 0.001$ to $-0.053 \pm 0.001$ (standard errors of the median, applied similarly in the following sections) once we account for daily mobility in the analysis. 
The reduced segregation reveals that, on average, native-born individuals are exposed to more social diversity in their daily activities, where they experience less segregation than in their residential areas.
Conversely, for the foreign-born individuals (Group F), the impact of mobility on segregation is limited, resulting in a modest reduction with average values shifting from $-0.380 \pm 0.002$ to $-0.295 \pm 0.001$. 
In general, mobility significantly alters individual segregation patterns, but this is not the case for foreign-born individuals in Sweden: foreign-born segregated residents largely remain segregated in their outside-home activities (see more details in SI Appendix, Residential vs. experienced segregation). \par

\subsection{Two Key Mechanisms: Destination Preference and Mobility Range}
Having revealed large differences between the segregation experience of native and foreign-born groups, it is crucial to understand why foreign-born individuals experience higher segregation during their daily activities. 
In this section, we assess the potential contributions of lifestyle, homophily, and mobility range through data exploration by comparing the mobility of foreign-born and native-born individuals.
We do not examine the role of urban structure, which sets a pre-condition for daily activities but can not be altered by individual decisions. \par

\begin{figure}[!tbhp]
\centering
\includegraphics[width=0.4\textwidth]{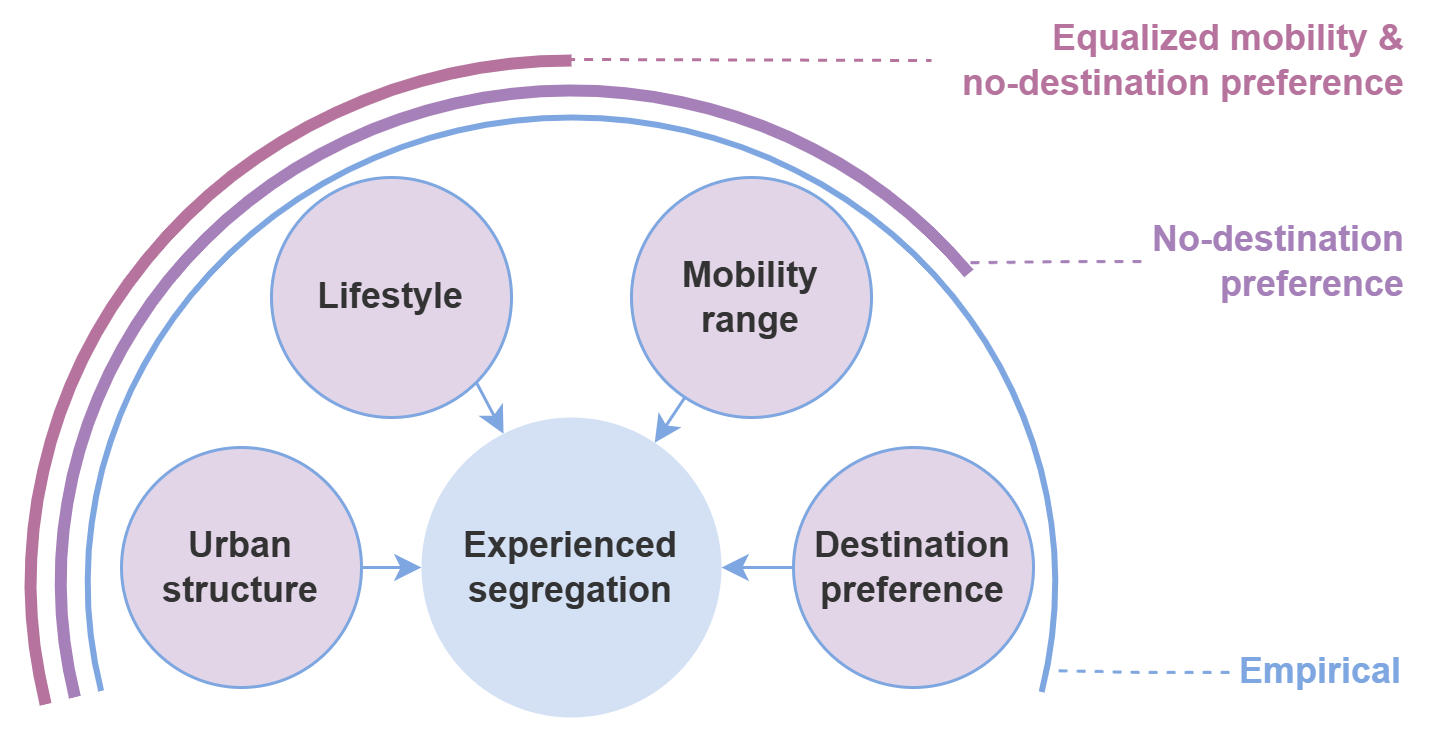}
\caption{\textbf{Key factors driving experienced segregation}. Empirically experienced segregation is impacted by urban structure, individuals' lifestyle, mobility range, and destination preference. The first counterfactual scenario minimizes the effect of destination preference, and the second equalizes mobility range across individuals.}
\label{fig:theories}
\end{figure}

Focusing first on lifestyle, we find that native-born (N) and foreign-born (F) individuals have similar activity levels and similar distribution patterns across categories of locations (see Fig. \ref{fig:seg_disp_fig1}c and SI Appendix, Fig. S7).
The differences measured by Cohen's d between Group F and Group N across all nine categories of locations indicate negligible disparities, ranging from -0.08 to 0.14. 
This suggests that \textit{lifestyle} choices have a limited impact on segregation in our dataset \citep{becker2000effect}. \par

Interestingly, we observe that within each category, individuals are more likely to visit locations frequently visited by others from their own group (see Figure \ref{fig:seg_disp_fig1}d). 
One hypothesis is that this pattern could be due to homophily, where individuals preferentially interact with those with similar characteristics or group affiliations.
A second hypothesis is that individuals tend to visit locations close to one's residence.  \par

Supporting the latter hypothesis, we observe correlations between individuals \textit{mobility range} and their experienced segregation.
Notably, compared to foreign-born individuals (Group F), native-born individuals (Group N) tend to travel further from home to reach their day-to-day destinations (see Fig. \ref{fig:seg_disp_fig1}e). 
We also find that, within each group, individuals' level of experienced segregation is associated with their radius of gyration \citep{xu2018human}, capturing the extent of their whereabouts (see Fig. \ref{fig:seg_disp_fig1}e). 
Individuals with larger mobility ranges tend to experience lower levels of segregation, especially for Group F (corr. coefficient = -0.21, $p<0.001$). \par

The findings above suggest that destination preference and mobility range could play a key role in experienced segregation. 
These two factors are, however, deeply intertwined, requiring further analysis to quantify their individual effects. \par  

\subsection{The Marginal Role of Destination Preference}
We quantify the effect of destination preference and mobility range on experienced segregation through counterfactual analysis based on simulations.
Initially, we test the hypothesis $H_h$ that experienced segregation in Group F is fully explained by destination preference. 
To this end, we perform simulations that maintain the real distribution of amenities and residences, lifestyle choices, and mobility ranges but where visit patterns are altered. 
In this \emph{no-destination preference} simulation, for each recorded trip in our dataset, we substitute the original destination with a random nearby location (within 1 km) of the same type (e.g., bar, restaurant, café, see more details in the Methods section).
To confirm or reject $H_h$, we measure $ICE_e$ in the counterfactual scenario, \emph{no-destination preference}. 
Significant values of $ICE_e$ would lead to the rejection of our hypothesis. 
To assess the significance of $ICE_e$ we perform significance tests against a Random mixing scenario (see Methods, Identifying segregated individuals). \par

Indeed, our analysis leads us to reject the hypothesis $H_h$.
Under the \emph{no-destination preference} scenario, we observe an average reduction in absolute experienced segregation $|ICE_e|$ for $71\%$ of the individuals in Group F.
Interestingly, while this reduction is statistically significant, its magnitude suggests having a small effect size according to Cohen's test (see Table \ref{tab:sims_effect}).
Under the \emph{no-destination preference} scenario, in fact, we still observe significant segregation for Group F, characterized by $ICE_e= -0.234 \pm 0.001$ (Fig. \ref{fig:seg_disp_fig2} and SI Appendix, Table S6).
For comparison, under the \emph{no-destination preference} scenario, Group N also experiences a small reduction in absolute experienced segregation ($58\%$ of individuals) and continues to present no significant segregation with $ICE_e= -0.061 \pm 0.001$ (SI Appendix, Table S6).
Overall, the analysis reveals that homophily plays a small but relevant role in explaining experienced segregation. 
In other words, when presented with two similar venues in close proximity, individuals from both Groups N and F exhibit a slight preference for visiting venues frequented by people similar to themselves.
This tendency is marginally more pronounced among Group F. 
However, this effect alone does not fully explain the high degree of experienced segregation observed in empirical data, with statistical test results detailed in SI Appendix, Section Statistical test results (Table S6 and Fig. S9). \par

\subsection{The Key Role of Mobility Range}
We proceed to test the hypothesis $H_{ht}$ that \emph{destination preference} and \emph{mobility range}, when combined, can explain the observed levels of segregation in Group F. \par

To evaluate this hypothesis, we conduct an \textit{equalized mobility \& no-destination preference} randomization simulation aimed at standardizing individuals' mobility range and minimizing their destination preferences (detailed in the Methods section). 
For each trip made by an individual in the dataset, we replace the destination with a random location of the same type located at a distance $d$ (with a buffer of $\pm$ 30 m) from the individual's home. 
To ensure that all individuals have the same mobility range, we extract the distance $d$ as a probability distribution from the average empirical distribution of distances between individuals' homes and visited locations (see the Methods section).
As in the previous section, to confirm or reject $H_{ht}$, we measure $ICE_e$ in the counterfactual scenario, where significant values of $ICE_e$ change would lead to rejecting $H_{ht}$. \par

\begin{figure}[!tbhp]
\centering
\includegraphics[width=0.5\linewidth]{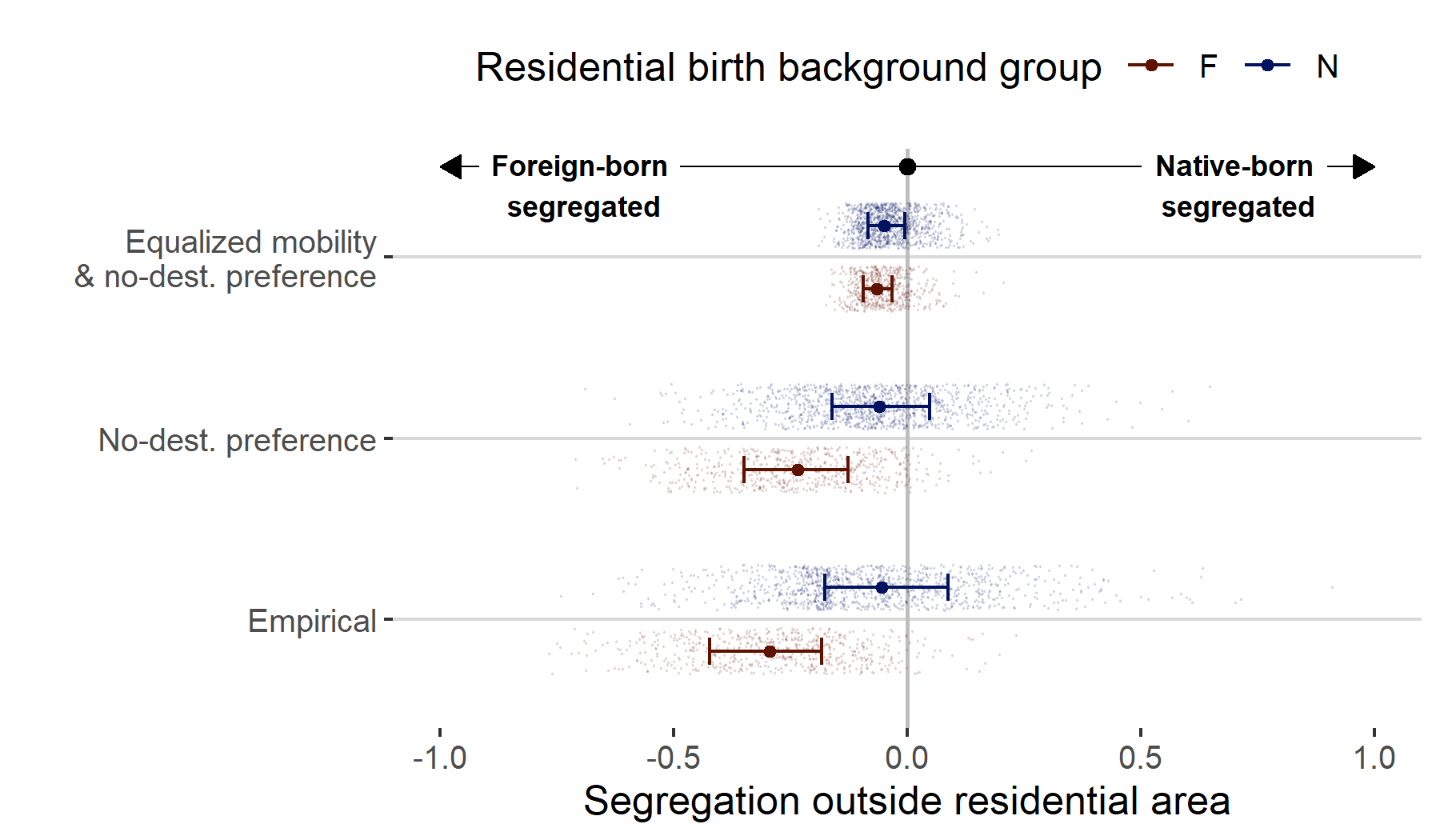}
\caption{\textbf{The effects of destination preference and mobility range on experienced segregation}.
Segregation ($ICE_e$) experienced by foreign-born (red dots) and native-born (blue dots) individuals in the empirical (bottom), first scenario (mid) and second scenario (top) conditions. Lines show the ranges between the 25th and 75th percentiles.}
\label{fig:seg_disp_fig2}
\end{figure}

We find that, under the counterfactual scenario, the average segregation for Group F is significantly reduced (see Table \ref{tab:sims_effect}) to a non-significant level of $ICE_e=-0.066\pm 0.0003$ (Fig. \ref{fig:seg_disp_fig2} and SI Appendix, Table S6). 
Hence, we cannot reject the hypothesis $H_{ht}$. 
Our result suggests that when combined, mobility range and destination preference can explain most of the level of experienced segregation observed in Group F.  
For comparison, under this counterfactual scenario, Group N also experiences a small reduction in absolute segregation, and continues to show no significant levels of segregation with $ICE_e= -0.049 \pm 0.0003$ (SI Appendix, Table S6).
Overall, the analysis reveals that limited mobility range plays a key role in the experienced segregation of foreign-born individuals. 
The corresponding statistical test results are detailed in SI Appendix, Statistical test results (Table S6 and Fig. S9).

\begin{table*}[t!]
\centering
\caption{\textbf{The effects of destination preference and mobility range on segregation}. The table shows the results of the statistical tests measuring the difference between empirical data and the two counterfactual scenarios for the native-born (N) and foreign-born (F) segregated groups. Diff. is the statistic of the Weighted Mann-Whitney U test between simulated and empirical distributions. All values are statistically significant, with $p<0.001$. The effect size is estimated based on Cohen's d, the magnitude of the difference between the two distributions. A Cohen's d value around 0.2 is typically considered a small effect. A value around 0.5 is considered a medium effect. A value around 0.8 or larger is considered a large effect \citep{becker2000effect}.}\label{tab:sims_effect}
\begin{tabular}{llrrr}
Effect                          & Group & Diff. & Cohen's d & Effect size \\\midrule
\multirow{2}{*}{Destination preference}      & N     & 0.012 & -0.06     & Very small  \\
                                & F     & -0.10 & 0.34      & Small       \\
\multirow{2}{*}{Destination preference \& mobility range} & N     & -0.02 & -0.01     & Very small  \\
                                & F     & -0.41 & 1.81      & Large \\
\bottomrule
\end{tabular}

\end{table*}

\subsection{Mobility Range and Destination Preference Shape Exposure between Groups}
\begin{figure*}[!tbhp]
\centering
\includegraphics[width=1\textwidth]{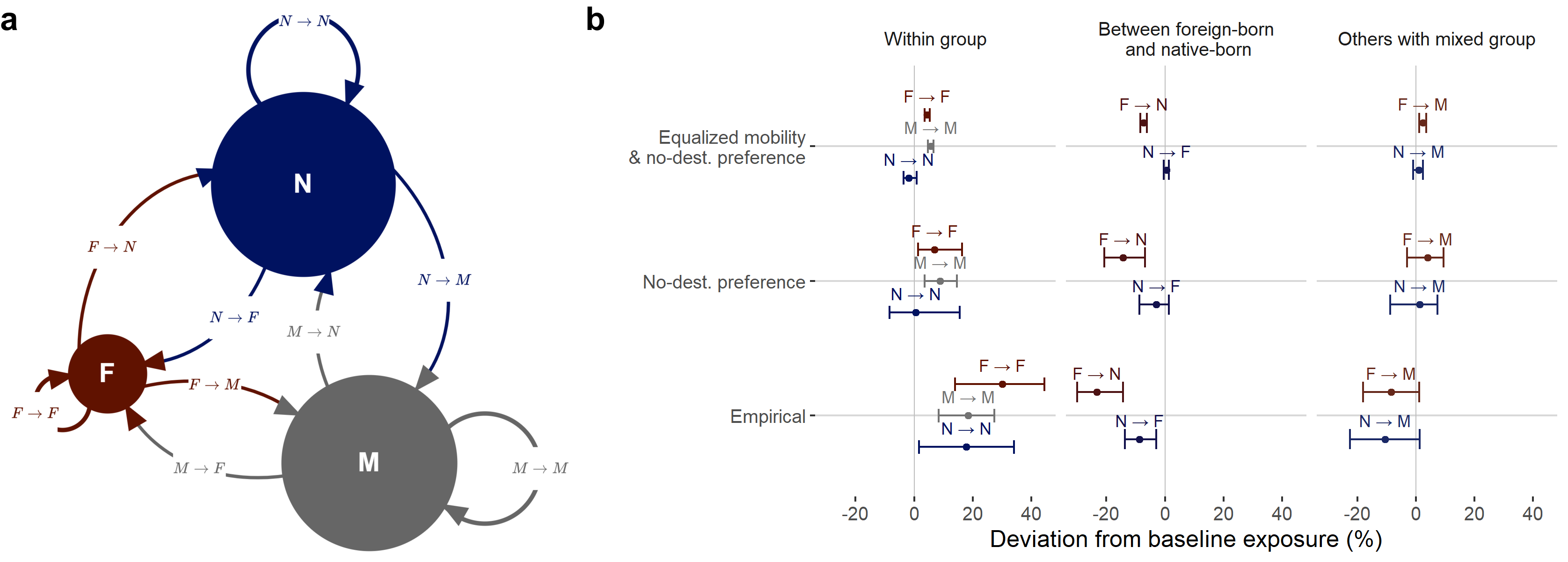}
\caption{\textbf{The effects of destination preference and mobility range on group exposure}.
\textbf{a}, Schematic network displaying exposure among individuals in foreign-born concentrated areas (Group F, red), native-born concentrated areas (Group N, blue), and mixed group (Group M, gray). Nodes correspond to groups, and their radii are proportional to the share of individuals in the data. The notation  $X \rightarrow Y$  indicates the typical fraction of an individual's encounters with individuals from Group $Y$, given that the individual is from Group $X$. \textbf{b}, Group exposure (see subplot c for the notation) in the empirical (bottom), first scenario (mid), and second scenario (top) conditions. Values are shown as the deviation from a baseline capturing expected exposure under random mixing conditions: $42.3\%$, $17.6\%$, and $40.1\%$ for exposure with Groups N, F, and M, respectively. Lines represent the ranges between the 25th and 75th percentiles.}
\label{fig:seg_groups}
\end{figure*}

Having established how destination preference and mobility range influence experienced segregation, we now look at how individuals from different groups are exposed to one another during their daily activities. 
This approach helps clarify the distinct contributions each group makes to experienced segregation, as shaped by these two mechanisms.
Specifically, we calculate the proportion of potential encounters each individual has with members of Groups N, F, and M (Fig. \ref{fig:seg_groups}a). 
In a scenario of random mixing, these proportions would reflect the distribution of individuals in each group: 42.3\% in Group N, 17.6\% in Group F, and 40.1\% in Group M.
Therefore, for each individual, we compute the deviation from these baseline values (Fig. \ref{fig:seg_groups}b).
The significance of such deviations is assessed by performing significance tests against a Random mixing scenario (see Methods, Identifying segregated individuals). 
In our notation, $X \rightarrow Y$ indicates the typical fraction of encounters an individual from Group $X$ has with individuals from Group $Y$. \par

Our findings reveal that individuals tend to be more exposed to others from their own group than expected. 
This effect is significant for individuals in Group F (detailed in SI Appendix, Statistical test results, Table S7 and Fig. S11).
Meanwhile, the other groups only present a slight tendency to have inner-group exposure that is not significant.
The proportions of $F \rightarrow F$, $N \rightarrow N$, and $M \rightarrow M$ encounters are $30.0 \pm 0.17\%$, $17.7 \pm 0.14\%$, and $18.4 \pm 0.06\%$ higher, respectively, than what would be expected under random mixing conditions. 
Conversely, exposure across groups, $F \rightarrow N$ and $N \rightarrow F$ are $-23.3 \pm 0.08\%$ and $-8.7 \pm 0.05\%$ lower than expected. \par

In the \emph{no-destination preference} scenario, Group N experiences encounters consistent with baseline (SI Appendix, Table S7), with $N \rightarrow N$ being only $0.6 \pm 0.1\%$ higher and $N \rightarrow F$ only $-3.1 \pm 0.02\%$ lower than expected.
Group F continues to experience a slightly unbalanced set of encounters, with $F \rightarrow F$ being $6.9 \pm 0.07\%$ higher and $F \rightarrow N$ being $-14.3 \pm 0.08\%$ lower than expected.
We note, however, that this result is not statistically significant (SI Appendix, Table S7), in apparent discrepancy with the results obtained in the previous section for the $ICE_e$.
This difference can be attributed to two factors. 
First, the categorical group exposure considers only three groups of individuals, providing less detail than the continuous quantification of the experienced segregation level.
Second, although Group F's exposures to Group F and Group N are insignificantly different from random mixing in the \textit{no-destination preference} scenario, the combined deviations result in Group F remaining largely segregated overall. \par

Finally, in the \textit{equalized mobility \& no-destination preference} scenario, Group N remains close to the baseline exposure to its own group ($N \rightarrow N$, $-1.8 \pm 0.02\%$) and Group F ($N \rightarrow F$, $0.6 \pm 0.01\%$). 
Similarly, in this scenario, the level of interaction of individuals in Group F with members of their own group ($F \rightarrow F$, $4.3 \pm 0.01\%$), and with Group N ($F \rightarrow N$, $-7.3 \pm 0.01\%$) do not deviate significantly from the baseline (see SI Appendix, Table S7). \par

In summary, these results show that destination preference has a minor but significant influence on group exposure of the foreign-born segregated population.
After minimizing the destination preference, there is no significant deviation from the baseline for Group F.
Further, equalizing the mobility range slightly narrows the gaps between Group F and the others.
For their native-born counterparts, these two factors have a minimal effect due to the already low level of experienced segregation in their daily activities outside the home. \par

\subsection{Reducing Segregation by Facilitating Mobility}
\begin{figure*}[!ht]
\centering
\includegraphics[width=1\textwidth]{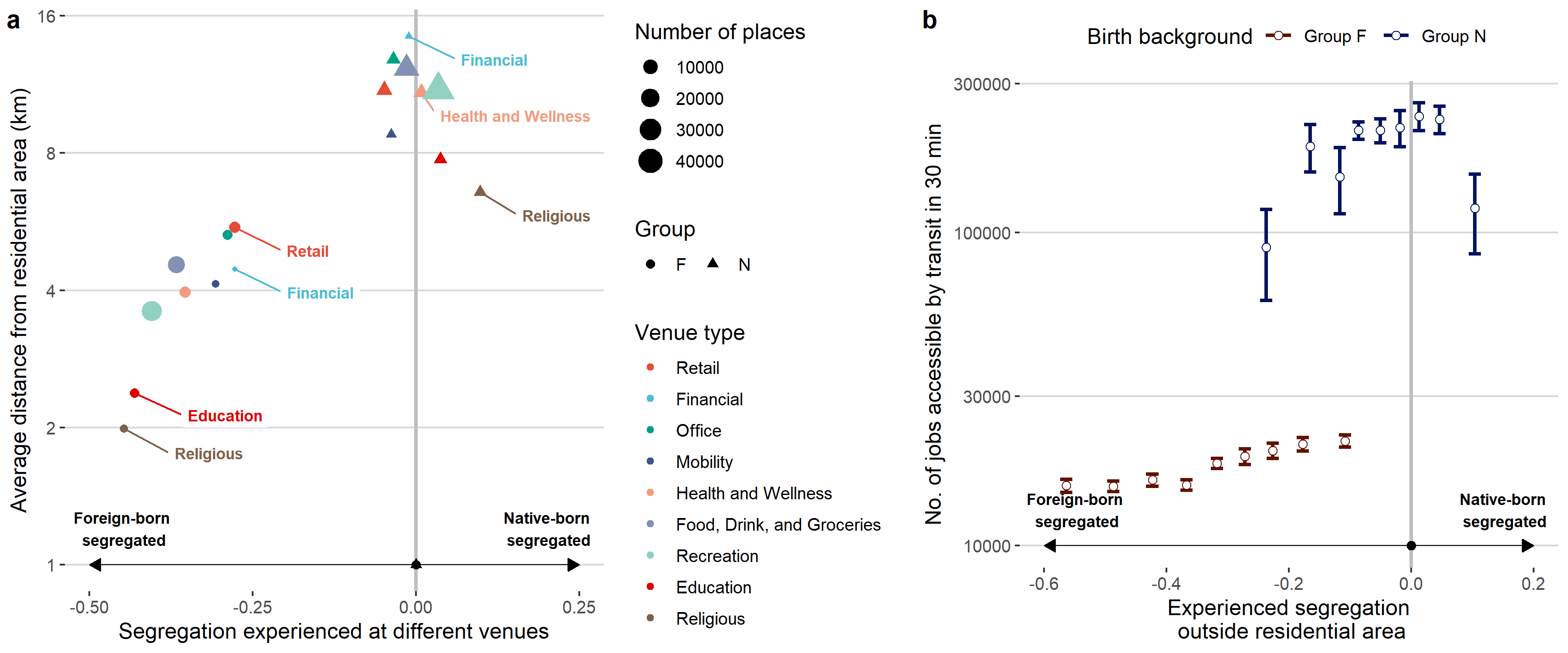}
\caption{\textbf{Segregation is associated with low mobility range and transport accessibility}. \textbf{a}, The average segregation experienced in visited destinations of different types against their average distance from visitors' homes. Results are shown for foreign-born (circles) and native-born (triangles) segregated individuals. \textbf{b}, Transit accessibility of individuals with low car ownership by their experienced segregation, divided into nine groups based on the quantile breaks of their respective segregation level. Transit accessibility is measured by the count of jobs reachable within 30 minutes by public transport from one's home. Low car ownership is identified as ownership rates under the 25th percentile, i.e., below 0.28 per capita.}\label{fig:seg_disp_fig3}
\end{figure*}

The results in the previous sections suggest that mobility range is a key factor driving differences in the experienced segregation of foreign-born and native-born individuals.
For both groups, the experienced segregation is higher in destinations closer to home, such as education or religious locations (see Fig. \ref{fig:seg_disp_fig3}a). 
Instead, experienced segregation is lower in destinations farther from home, such as retail and financial locations (see Fig. \ref{fig:seg_disp_fig3}a). \par

To understand the role of transport accessibility in segregation, we investigate their relationship by focusing on individuals with low car ownership. 
This specific group is assumed to rely primarily on public transportation and active modes of travel.
We observe that, within this group, individuals in Group F have much lower job accessibility by transit than those in Group N (see Fig. \ref{fig:seg_disp_fig3}c).
In Group F, higher job accessibility by public transit is associated with less experienced segregation outside residential areas (corr. coefficient = -0.16, $p<0.001$). \par

Our results reveal that longer mobility ranges and better transport accessibility are associated with reduced segregation. \par

\section{Discussion}
In this study, we used smartphone GPS records to examine the division between native-born and foreign-born (outside Europe) populations in Sweden.
We found that the segregation experiences of foreign-born and native-born individuals are affected differently by their out-of-home activities.
Native-born individuals experience non-significant levels of segregation while performing routine activities outside the home.
In contrast, residents of foreign-born segregated areas experience less, but still significant, levels of segregation during their outside-home routine activities because they tend to visit locations popular with other foreign-born individuals. 
Through counterfactual simulations, we tested two hypotheses to understand the factors driving this heterogeneous effect. 
First, we explored whether foreign-born individuals prefer destinations frequented by those with similar backgrounds (homophily hypothesis).
We observed that destination preference is only marginally responsible for the segregation levels experienced by foreign-born individuals. 
Secondly, we tested whether the observed heterogeneous effect of mobility patterns on segregation results due to differences in mobility ranges between foreign-born and native-born individuals. 
Our findings indicate that mobility constraints play a crucial role, with foreign-born individuals traveling shorter distances around their residential areas compared to native-born. 
Through counterfactual simulations, we demonstrated that if everyone had the same mobility range and no destination preference, the segregation experienced by foreign-born individuals would be negligible. 
Importantly, we found that limited mobility ranges among foreign-born individuals are associated with lower car ownership and reduced transport accessibility. \par

Our analysis contributes to the recent stream of work on mobility and segregation by analyzing GPS trajectories collected from smartphones \citep{liao2024socio}.
Our findings suggest that, in Sweden, day-to-day activities overall show a lower level of segregation by country of origin than the residential level, similar to what's revealed in several previous studies \citep{park2018beyond, garlick2022there, fuentes2022impact, athey2021estimating, moya2021exploring, shin2021spatial, silm2021relationship, xian2022beyond, xu2022beyond, kronenberger2017configurational, atuesta2018access}. 
They align with previous studies, showing that mobility range relates to individuals' segregation experiences and that activities/locations close to individual residences are more segregated \citep{moro2021mobility}. 
Additionally, visits to venues in the city centre, such as retail locations \citep{nilforoshan2023human}, are associated with lower levels of experienced segregation. 
Previous studies showed that traveling outside one's residence generally reduces the level of experienced segregation \citep{osth2018spatial,abbiasov202415}. \par

We acknowledge some limitations in our analysis. 
First, similar to previous research \citep{moro2021mobility,nilforoshan2023human}, we estimate individuals' socioeconomic attributes based on census data from their areas of residence. 
Using data that integrates both socioeconomic attributes and mobility information would improve the accuracy of our analysis. 
Secondly, our counterfactual scenarios treat broad categories of locations (e.g., restaurants, cafés) as homogenous, overlooking differences in cost, attractiveness, and other characteristics. 
Third, the observed relationship between transport access and segregation is correlational. 
Future studies using longitudinal data could explore whether improved transport access facilitates greater interaction between foreign-born and native-born individuals. 
Finally, mere co-presence or exposure in a given area does not always equate to meaningful social interactions \citep{schnell2014arab,zhou2019between,dorman2020does}, despite co-presence being a precursor to social interaction \citep{collins2004interaction,netto2015segregated}. 
To guide policymaking that promotes social integration among different groups, further research is needed on how to accurately capture and represent their actual interactions. \par

Overall, our work suggests that improving transport accessibility could help reduce social segregation \citep{huang2022unfolding}.
Enhanced accessibility broadens individuals' freedom to reach opportunities across employment, healthcare, and education services \citep{pereira2017distributive}.  
Prioritizing accessible and affordable transport solutions, particularly in areas with high foreign-born populations, could promote social mixing and create more inclusive communities. \par

\section{Methods}\label{sec:methods}
\subsection{Data Preparation and Segregation Computation}\label{sec:data_seg}
This study uses anonymized mobile phone data from GPS records collected through location-enabled smartphone applications. 
The dataset includes individuals who consented to anonymously share their data through a General Data Protection Regulation-compliant framework.
In 2020, data was shared under a contract with Pickwell, who provided access to de-identified and privacy-enhanced mobility data. 
All researchers were required not to share the data further or attempt to re-identify it. \par

The dataset covers seven months in 2019 (June-Dec) with about 25 million daily GPS time records from 1 million devices residing in Sweden, giving a total of approximately 5.2 billion records, i.e., 5,250 records per device on average.
Assuming that a device is a person, the population covered by this dataset is equivalent to c.a. 10\% of the Swedish residents. 
After further processing, we focus on individuals who reside in Sweden, have sufficient data records, and have reliably identified home zones (see SI Appendix, Section Data filtering).
More details of the dataset can be found in SI Appendix, section 'Mobility data'. \par

\subsection{Stay and Home Detection}\label{sec:mobi_proc}
To measure experienced social segregation, we need to measure when and where people are engaged in activities. 
To do so, the first step is to detect \textit{stays} from individual geolocation records (\textit{id, lat, lon, time}).
A stay is defined as an instance in which an individual spends a significant amount of time within the same area.  
These \textit{stay} observations are used for further analysis to characterize individuals' travel and mobility segregation patterns. 
There are several algorithms for detecting stays \citep{aslak2020infostop, hariharan2004project}. 
In this study, we use the Infostop algorithm \citep{aslak2020infostop} due to its robustness against measurement noise, scalability to large datasets, and capability of doing multi-user analysis simultaneously. 
The details, such as the selection of the algorithm parameters, are specified in SI Appendix, section 'Stay detection'. \par

Once stays are detected, we identify individuals' residential areas to match them with socioeconomic attributes from census data.
Residential areas are population grids or census zones that correspond to the most probable location of residence for an individual device. 
Following previous studies \citep{moro2021mobility}, we identify residences by first identifying the top 3 most visited places during non-holiday time\footnote{Holiday time: 2019-06-23--2019-08-11 and 2019-12-22--2020-01-01. 
These times are when most people take summer and winter vacations in Sweden.}, and then select the most visited place between 10:00 p.m. and 6:00 a.m. 
In this step, we filter out those individuals without reliable home areas or sufficient data.
More details can be found in SI Appendix, section 'Data filtering'. \par

\subsection{Socioeconomic Attributes Assignment}\label{sec:census}
We collect census data from the statistics agency in Sweden \citep{SCB-deso}. 
Data come with two systems of spatial units: Demographic statistics areas (DeSO zones) and population grids with sizes of 250 m or 1000 m (SI Appendix, Section Census data).
Census data includes information on the following characteristics of each spatial unit (grid and DeSO zone) that are used in this study: total population, birth background (Sweden, outside Sweden in Europe, outside Europe, and others), and job count (daytime job register).
These attributes are assigned to each individual device according to the census data and where a given device resides. 
In this study, we use DeSo zones to calculate residential segregation, and population grids to summarize individuals' experienced segregation and for transport accessibility computation.
SI Appendix, section 'Census data' contains more details of the data used.

\subsection{Reducing Sampling and Population Biases}\label{sec:debias}
Mobile phone application data records the population's geolocations when they use certain applications.
Therefore, we have more recorded locations when people are more likely to use their phones, i.e., in the afternoon, evening, and night, compared with other times.
We design stay weight ($W_{pr}$ for record $r$ of individual $p$) to reduce the sampling bias introduced by imbalanced knowledge of individuals' geolocations due to this passive data collection. 
Specifically, we assign greater significance to time intervals during which fewer locations are registered and lesser significance to those with more recorded locations.
This approach helps reduce the sampling bias toward activities at certain times of the day when we have more records. 
For more details on the stay weight design, please see SI Appendix, section 'Sampling debiasing'. \par

Another bias of mobility data stems from who uses these mobile devices, i.e., how representative the phone users are compared with the actual population.
This study assigns weights to individual mobile devices based on their home zones. 
It uses Inverse Probability Weighting (IPW) to give more weight to less densely populated areas, reducing bias from high device concentration. 
Extreme weights in sparsely populated areas are controlled using weight trimming. 
This ensures balanced weights, enhancing the accuracy and reliability of forthcoming statistical analyses.
The created individual weight has a notation $W_p$ for a given individual $p$.
The weight reduces the population bias, where we have proportionally more devices in big cities than in the other parts of Sweden, providing a better representation of the population. 
SI Appendix, section 'Population debiasing' contains more details of the population weight design. \par

At the end of this process, we have a data set of stay records representative of individuals residing in Sweden, which can be used to calculate their social segregation levels. \par

\subsection{Segregation Indicator}\label{sec:seg_ind}
We measure the social segregation between foreign-born (outside of Europe) and native-born populations using a modified version of the Indicator of Concentration at Extremes (ICE) \citep{massey2001residential}, which can be calculated using Equation \ref{eq:ice} for a given zone $j$. 
This indicator measures how the areas deviate from their national average composition of birth background, which offers a comprehensive and actionable understanding of segregation between native-born and foreign-born. 
It places local segregation patterns in a national context, aiding in informed decision-making and policy development. \par

\begin{equation}
ICE_j = \frac{\frac{N_j}{w_N} - \frac{F_j}{w_F}}{\frac{N_j}{w_N} + \frac{F_j}{w_F} + \frac{O_j}{w_O}} \label{eq:ice}
\end{equation}%
Here, $ F_j $ is the number of foreign-born (outside of Europe) individuals, $ N_j $ the number of native-born individuals, and $ O_j $ the remaining individuals.
The overall fraction of native-born, foreign-born outside of Europe and other populations at the national level are $ w_N $, $ w_F $, and $ w_O $, respectively. 
We use the data from 2019 when $ w_N = 0.804 $, $ w_F = 0.111 $, and $ w_O = 0.085 $. 
It is worth noting that the original formula of $ICE$ does not include population weights. 
We adopt the adjusted $ICE$ using weights because we focus on the relative value of birth background segregation. 
The comparison between our indicator and the original one is detailed in SI Appendix, section Adjusted ICE vs. its original form. \par

$ ICE_j $ ranges between -1 and 1. 
In a location with $ ICE_j = 0 $, the composition of subpopulations equals the national average, $ ICE_j = 1 $ means that there are only native-born living (or being active) in zone $ j $ and $ ICE_j = -1 $ indicates only foreign-born (outside of Europe). \par

\subsection{Residential and Experienced Segregation}\label{sec:exp_seg}
\textbf{Residential segregation} $ICE_r$ is calculated using Census data following Equation \ref{eq:ice}, with $ F_j $ and $ N_j $ the foreign-born (outside of Europe) and native-born residents in a census zone $ j $. \par

Individuals' mobility time history tells us the places they visit, when, and for how long, and the segregation levels at those places describe the individually experienced segregation.
To calculate \textbf{experienced segregation}, we need to know the visiting segregation of all the analysis zones (detailed in SI Appendix, Section Analysis zones) an individual visited and aggregate them. 
Therefore, we first quantify the visiting segregation for various zones across different times.
We use the mobility data to find all individuals co-present in each spatiotemporal unit, i.e., which analysis zone (see SI Appendix, Section Analysis zones) and time interval, where a day is divided into 48 half-hour intervals starting from 00:00 ($i$=1,2,...,48).
We also define whether a day is a weekday or a weekend/holiday.
As a result, we have $2 \times 2 \times 48=192$ temporal units, considering 48 time intervals of a day, weekday/weekend, and holiday/non-holiday. \par

Visiting segregation of a given analysis zone $ j $ at a temporal unit $\left(i,weekday,holiday\right)$ is calculated using the co-present visitors following Equation \ref{eq:ice}.
Unlike the residential segregation measuring residents, we need to calculate $ F $ and $ N $ for visitors.
Each visitor has these attributes from their residential population grid's statistics (Census data).
We calculate the weighted sum of these attributes using their population weight to get a more representative picture of the composition of the co-present visitors and, therefore, the level of visiting segregation $ICE_v\left(i,j,weekday,holiday\right)$. \par

This study continues with the aggregate picture of individually experienced segregation $ICE_e$ in daily mobility to simplify the analysis ($weekday=1$, $holiday=0$).
For each time interval $i$, an individual may have stay records in a few different zones $j$.
Therefore, we use the median value of $ICE_v\left(j\right)$ to represent the average experienced segregation at a given temporal unit $ICE_e\left(i\right)$.
We further simplify the analysis by calculating $ICE_e$ for an average non-holiday weekday as the mean value of $ICE_e\left(i\right),i=1,2,...,48$. \par

At the end of this calculation, each individual has one residential segregation level $ICE_r$ depending on where they reside and one experienced segregation level $ICE_e$ derived from their mobility outside residential areas.

\subsection{Identifying Segregated Individuals}\label{sec:seg_sig}
Based on the residential segregation level of individuals, we create three groups: those segregated towards foreign-born (Group F), those segregated towards native-born (Group N), and the rest with insignificant segregation (Group M).
We conduct a residential randomization simulation to establish thresholds for identifying segregated individuals. 
This simulation emulates a scenario where individuals in Sweden make random housing/visiting choices, leading to a diverse mix of foreign-born, native-born, and others.
We call this scenario Random mixing, which serves as the comparison baseline.
In this process, we randomly reassign residences among individuals and then assess the level of residential segregation for each person.
After completing this randomization process 100 times, we compile and analyze the distribution of residential segregation levels for these individuals. \par

Residential segregation values falling outside the 99\% confidence interval are deemed statistically segregated. 
Consequently, thresholds of -0.2 and 0.2 are established for classifying individuals. 
Specifically, those with a residential segregation level below -0.2 are categorized into Group F (foreign-born concentrated), those above 0.2 into Group N (native-born concentrated), and those between -0.2 and 0.2 are placed in Group M (insignificantly segregated). \par

Each individual is also categorized into one of the same three groups based on their experienced segregation level sequence at each half-hour interval during an average weekday, $ICE_e\left(i\right),i=1,2,...,48$. 
If the sequence is significantly greater than 0.2, the individual is classified into Group N. 
Conversely, if the sequence is significantly less than -0.2, they are placed in Group F. 
In all other scenarios, the individual falls into Group M. 
For sequences forming a normal distribution, we apply a one-sample t-test; for non-normal distributions, we utilize the Wilcoxon test. \par

In the Random mixing scenario, each individual is exposed to others outside their home with the following composition: 42.3\% from Group N, 17.6\% from Group F, and 40.1\% from Group M, reflecting the distribution of these groups in our dataset. 
To determine significant deviations from this random mixing, we design a procedure to obtain confidence intervals for the share of encounters outside the home. 
Following the randomization process, we identify the 95\% confidence intervals for the presence of the three groups in each individual's encounters. 
Specifically, any share outside the ranges (-19\%, 19\%) for Groups N and M, or (-13\%, 13\%) for Group F, is considered a significant deviation from random mixing. 
For the group exposure distribution, we utilize the Wilcoxon test to see if it falls outside its 95\% confidence interval.
If the group exposure distribution is significantly greater than the upper bound or less than the lower bound, it significantly deviates from the Random mixing exposure. 
In all other scenarios, there is no significant deviation from the Random mixing. 
The details of the above statistical results can be found in SI Appendix, section 'Statistical test results'. \par

\subsection{No-destination Preference Simulation}\label{sec:pref_sim}
The destination preference hypothesis refers to individuals who tend to visit places where there are people similar to them.
The No-destination preference simulation aims to reduce the impact of individual preference on the visited places by randomly shifting them to nearby places with similar functions. \par

We first assign each non-home stay to a point of interest, i.e., the nearest POI within a search radius of 300 m.
Then, we randomly shift the non-home stays to a nearby POI with the same category or similar kind.
This shifted stay is randomly selected from the POIs between the search radii of 30 m and 1000 m around the original stay record. 
The details of POI data and how the random shifting is implemented are specified in SI Appendix, section 'Point of interest data'. \par

For each individual, we ran this simulation 50 times.
Finally, we repeat the same calculation described in section 'Residential and Experienced Segregation' to get the simulated values of experienced segregation.

\subsection{Equalized Mobility \& Mo-destination Preference Simulation}\label{sec:limi_sim}
We hypothesize that individuals who are reluctant or face challenges in traveling contribute to restricted social mixing with other population groups due to their limited mobility range.
To quantify the contribution of mobility range in segregation experiences, we design the Equalized mobility \& no-destination preference simulation that equalizes travel distance disparities among individuals in our dataset by assuming a uniform mobility range across them. \par

We begin by documenting all the locations individuals visited and calculating the distances from their home locations.
Focusing on distances less than 1,500 km, we establish bins with 1 km intervals to compute the frequency of visits within each distance range. 
The resulting distributions, illustrated in SI Appendix, Fig. S2, encompass all individuals, those residing in areas with foreign-born populations and those with native-born populations. 
This simulation assumes one travels as an average person, following the distance-decay trend (the gray distribution in SI Appendix, Fig. S2). 
In this case, the individuals in Group F will increase their mobility range, while Group N will slightly decrease their mobility range compared to their respective empirical mobility patterns. \par

\section*{Data availability}
The data supporting the findings of this study were purchased from PickWell and are subject to restrictions due to licensing and privacy considerations under the European General Data Protection Regulation. 
Consequently, these data are not publicly available, but are commercially available and may be requested for research use (\url{https://www.pickwell.co/}). 
Aggregated data to reproduce all results are publicly available at \url{https://github.com/MobiSegInsights/mobi-seg-se/tree/main/data}. 
Venue locations and categories can be retrieved from OpenStreetMap (\url{https://download.geofabrik.de}).
Census data (DeSO zones and their statistics) were collected from Statistics Sweden (\url{https://www.scb.se/}) that is publicly available. 
Census data (population grids) were collected from the Swedish University of Agricultural Sciences (\url{https://maps.slu.se/}) and were restricted to use only by individuals associated with Swedish research institutes.
GTFS data were collected from Samtrafiken API (\url{https://samtrafiken.se/}) that are publicly available. 
All data were utilized in accordance with the terms of service specified by their respective provider. \par

We strictly adhered to the guidelines set forth by the Chalmers Institutional Review Board (IRB) in accordance with the Swedish Act (2003:460) concerning the ethical review of research involving humans, as well as the General Data Protection Regulation 2016/679 (GDPR). 
Due to the nature of the data analysed, the study was exempt from ethical review under the Swedish Ethical Review Act (2003:460). \par

Python (version 3.10) code and R (version 4.0.2) code were used to analyse and visualize the data. 
The stays have been detected via infostop (version 0.1.11) and pyspark (version 3.5.1).
Code to reproduce our results is publicly available on GitHub https://github.com/MobiSegInsights/mobi-seg-se.

\bibliographystyle{unsrtnat}
\bibliography{references}

\begin{thebibliography}{54}
\providecommand{\natexlab}[1]{#1}
\providecommand{\url}[1]{\texttt{#1}}
\expandafter\ifx\csname urlstyle\endcsname\relax
  \providecommand{\doi}[1]{doi: #1}\else
  \providecommand{\doi}{doi: \begingroup \urlstyle{rm}\Url}\fi

\bibitem[Musterd et~al.(2017)Musterd, Marci{\'n}czak, Van~Ham, and Tammaru]{musterd2017socioeconomic}
Sako Musterd, Szymon Marci{\'n}czak, Maarten Van~Ham, and Tiit Tammaru.
\newblock Socioeconomic segregation in european capital cities. increasing separation between poor and rich.
\newblock \emph{Urban geography}, 38\penalty0 (7):\penalty0 1062--1083, 2017.

\bibitem[Chetty et~al.(2022)Chetty, Jackson, Kuchler, Stroebel, Hendren, Fluegge, Gong, Gonzalez, Grondin, Jacob, et~al.]{chetty2022social}
Raj Chetty, Matthew~O Jackson, Theresa Kuchler, Johannes Stroebel, Nathaniel Hendren, Robert~B Fluegge, Sara Gong, Federico Gonzalez, Armelle Grondin, Matthew Jacob, et~al.
\newblock Social capital i: measurement and associations with economic mobility.
\newblock \emph{Nature}, 608\penalty0 (7921):\penalty0 108--121, 2022.

\bibitem[Sousa and Nicosia(2022)]{sousa2022quantifying}
Sandro Sousa and Vincenzo Nicosia.
\newblock Quantifying ethnic segregation in cities through random walks.
\newblock \emph{Nature Communications}, 13\penalty0 (1):\penalty0 5809, 2022.

\bibitem[Iceland and Nelson(2010)]{iceland2010residential}
John Iceland and Kyle~Anne Nelson.
\newblock The residential segregation of mixed-nativity married couples.
\newblock \emph{Demography}, 47\penalty0 (4):\penalty0 869--893, 2010.

\bibitem[Malmberg et~al.(2013)Malmberg, Andersson, and {\"O}sth]{malmberg2013segregation}
Bo~Malmberg, Eva Andersson, and John {\"O}sth.
\newblock Segregation and urban unrest in sweden.
\newblock \emph{Urban geography}, 34\penalty0 (7):\penalty0 1031--1046, 2013.

\bibitem[Jarvis et~al.(2023)Jarvis, Mare, and Nordvik]{jarvis2023assortative}
Benjamin~F Jarvis, Robert~D Mare, and Monica~K Nordvik.
\newblock Assortative mating, residential choice, and ethnic segregation.
\newblock \emph{Research in Social Stratification and Mobility}, page 100809, 2023.

\bibitem[Algan et~al.(2012)Algan, Bisin, Manning, and Verdier]{algan2012cultural}
Yann Algan, Alberto Bisin, Alan Manning, and Thierry Verdier.
\newblock \emph{Cultural integration of immigrants in Europe}.
\newblock Oxford University Press, 2012.

\bibitem[Liao et~al.(2024)Liao, Gil, Yeh, Pereira, and Alessandretti]{liao2024socio}
Yuan Liao, Jorge Gil, Sonia Yeh, Rafael~HM Pereira, and Laura Alessandretti.
\newblock Socio-spatial segregation and human mobility: A review of empirical evidence.
\newblock \emph{arXiv preprint arXiv:2403.06641}, 2024.

\bibitem[Duncan and Duncan(1955)]{duncan1955methodological}
Otis~Dudley Duncan and Beverly Duncan.
\newblock A methodological analysis of segregation indexes.
\newblock \emph{American sociological review}, 20\penalty0 (2):\penalty0 210--217, 1955.

\bibitem[Feitosa et~al.(2007)Feitosa, Camara, Monteiro, Koschitzki, and Silva]{feitosa2007global}
Fl{\'a}via~F Feitosa, Gilberto Camara, Ant{\^o}nio Miguel~Vieira Monteiro, Thomas Koschitzki, and Marcelino~PS Silva.
\newblock Global and local spatial indices of urban segregation.
\newblock \emph{International Journal of Geographical Information Science}, 21\penalty0 (3):\penalty0 299--323, 2007.

\bibitem[Barros and Feitosa(2018)]{barros2018uneven}
Joana Barros and Flavia~F Feitosa.
\newblock Uneven geographies: Exploring the sensitivity of spatial indices of residential segregation.
\newblock \emph{Environment and Planning B: Urban Analytics and City Science}, 45\penalty0 (6):\penalty0 1073--1089, 2018.

\bibitem[Kwan(2013)]{kwan2013beyond}
Mei-Po Kwan.
\newblock Beyond space (as we knew it): Toward temporally integrated geographies of segregation, health, and accessibility: Space--time integration in geography and giscience.
\newblock \emph{Annals of the Association of American Geographers}, 103\penalty0 (5):\penalty0 1078--1086, 2013.

\bibitem[Moro et~al.(2021)Moro, Calacci, Dong, and Pentland]{moro2021mobility}
Esteban Moro, Dan Calacci, Xiaowen Dong, and Alex Pentland.
\newblock Mobility patterns are associated with experienced income segregation in large us cities.
\newblock \emph{Nature communications}, 12\penalty0 (1):\penalty0 4633, 2021.

\bibitem[Li et~al.(2022)Li, Yue, Gao, Zhong, and Barros]{li2022towards}
Qing-Quan Li, Yang Yue, Qi-Li Gao, Chen Zhong, and Joana Barros.
\newblock Towards a new paradigm for segregation measurement in an age of big data.
\newblock \emph{Urban Informatics}, 1\penalty0 (1):\penalty0 5, 2022.

\bibitem[Nilforoshan et~al.(2023)Nilforoshan, Looi, Pierson, Villanueva, Fishman, Chen, Sholar, Redbird, Grusky, and Leskovec]{nilforoshan2023human}
Hamed Nilforoshan, Wenli Looi, Emma Pierson, Blanca Villanueva, Nic Fishman, Yiling Chen, John Sholar, Beth Redbird, David Grusky, and Jure Leskovec.
\newblock Human mobility networks reveal increased segregation in large cities.
\newblock \emph{Nature}, pages 1--7, 2023.

\bibitem[Xu et~al.(2024)Xu, Wang, Attia, Attia, Zhang, and Zong]{xu2024experienced}
Wenfei Xu, Zhuojun Wang, Nada Attia, Youssef Attia, Yucheng Zhang, and Haotian Zong.
\newblock An experienced racial-ethnic diversity dataset in the united states using human mobility data.
\newblock \emph{Scientific Data}, 11\penalty0 (1):\penalty0 638, 2024.

\bibitem[Yang et~al.(2023)Yang, Pentland, and Moro]{yang2023identifying}
Yanni Yang, Alex Pentland, and Esteban Moro.
\newblock Identifying latent activity behaviors and lifestyles using mobility data to describe urban dynamics.
\newblock \emph{EPJ Data Science}, 12\penalty0 (1):\penalty0 15, 2023.

\bibitem[Xu et~al.(2022)Xu, Santi, and Ratti]{xu2022beyond}
Yang Xu, Paolo Santi, and Carlo Ratti.
\newblock Beyond distance decay: Discover homophily in spatially embedded social networks.
\newblock \emph{Annals of the American Association of Geographers}, 112\penalty0 (2):\penalty0 505--521, 2022.

\bibitem[Huang et~al.(2022)Huang, Zhao, Wang, Li, Yang, Feng, Xu, Zhu, and Chen]{huang2022unfolding}
Xiao Huang, Yuhui Zhao, Siqin Wang, Xiao Li, Di~Yang, Yu~Feng, Yang Xu, Liao Zhu, and Biyu Chen.
\newblock Unfolding community homophily in us metropolitans via human mobility.
\newblock \emph{Cities}, 129:\penalty0 103929, 2022.

\bibitem[Chen et~al.(2024)Chen, Zhou, Lu, and Zheng]{chen2024behavioral}
Fei Chen, Suhong Zhou, Junwen Lu, and Zhong Zheng.
\newblock A behavioral explanation of the activity-space segregation: Individuals’ preference of choosing an activity destination.
\newblock \emph{Environment and Planning B: Urban Analytics and City Science}, page 23998083241229110, 2024.

\bibitem[Vachuska(2023)]{vachuska2023racial}
Karl Vachuska.
\newblock Racial segregation in everyday mobility patterns: Disentangling the effect of travel time.
\newblock \emph{Socius}, 9:\penalty0 23780231231169261, 2023.

\bibitem[Abbiasov et~al.(2024)Abbiasov, Heine, Sabouri, Salazar-Miranda, Santi, Glaeser, and Ratti]{abbiasov202415}
Timur Abbiasov, Cate Heine, Sadegh Sabouri, Arianna Salazar-Miranda, Paolo Santi, Edward Glaeser, and Carlo Ratti.
\newblock The 15-minute city quantified using human mobility data.
\newblock \emph{Nature Human Behaviour}, pages 1--11, 2024.

\bibitem[Netto et~al.(2015)Netto, Soares, and Paschoalino]{netto2015segregated}
Vinicius~M Netto, Ma{\'\i}ra~Pinheiro Soares, and Roberto Paschoalino.
\newblock Segregated networks in the city.
\newblock \emph{International Journal of Urban and Regional Research}, 39\penalty0 (6):\penalty0 1084--1102, 2015.

\bibitem[Tao et~al.(2020)Tao, He, and Luo]{tao2020influence}
Sui Tao, Sylvia~Y He, and Shuli Luo.
\newblock The influence of job accessibility on local residential segregation of ethnic minorities: A study of hong kong.
\newblock \emph{Population, Space and Place}, 26\penalty0 (8):\penalty0 e2353, 2020.

\bibitem[Patias et~al.(2023)Patias, Rowe, and Arribas-Bel]{patias2023local}
Nikos Patias, Francisco Rowe, and Dani Arribas-Bel.
\newblock Local urban attributes defining ethnically segregated areas across english cities: A multilevel approach.
\newblock \emph{Cities}, 132:\penalty0 103967, 2023.

\bibitem[{OECD}(2022)]{oecd2022international}
{OECD}.
\newblock \emph{International migration outlook 2022}.
\newblock OECD, 2022.

\bibitem[Becker(2000)]{becker2000effect}
Lee~A Becker.
\newblock Effect size (es).
\newblock 2000.

\bibitem[Xu et~al.(2018)Xu, Belyi, Bojic, and Ratti]{xu2018human}
Yang Xu, Alexander Belyi, Iva Bojic, and Carlo Ratti.
\newblock Human mobility and socioeconomic status: Analysis of singapore and boston.
\newblock \emph{Computers, Environment and Urban Systems}, 72:\penalty0 51--67, 2018.

\bibitem[Park and Kwan(2018)]{park2018beyond}
Yoo~Min Park and Mei-Po Kwan.
\newblock Beyond residential segregation: A spatiotemporal approach to examining multi-contextual segregation.
\newblock \emph{Computers, Environment and Urban Systems}, 71:\penalty0 98--108, 2018.

\bibitem[Garlick et~al.(2022)Garlick, Catney, Darlington-Pollock, and Lloyd]{garlick2022there}
Sarah Garlick, Gemma Catney, Frances Darlington-Pollock, and Christopher~D Lloyd.
\newblock Is there greater ethnic mixing in residential or workplace spaces?
\newblock \emph{Journal of Ethnic and Migration Studies}, pages 1--21, 2022.

\bibitem[Fuentes et~al.(2022)Fuentes, Truffello, and Flores]{fuentes2022impact}
Luis Fuentes, Ricardo Truffello, and M{\'o}nica Flores.
\newblock Impact of land use diversity on daytime social segregation patterns in santiago de chile.
\newblock \emph{Buildings}, 12\penalty0 (2):\penalty0 149, 2022.

\bibitem[Athey et~al.(2021)Athey, Ferguson, Gentzkow, and Schmidt]{athey2021estimating}
Susan Athey, Billy Ferguson, Matthew Gentzkow, and Tobias Schmidt.
\newblock Estimating experienced racial segregation in us cities using large-scale gps data.
\newblock \emph{Proceedings of the National Academy of Sciences}, 118\penalty0 (46):\penalty0 e2026160118, 2021.

\bibitem[Moya-Gómez et~al.(2021)Moya-Gómez, Stępniak, García-Palomares, Frías-Martínez, and Gutiérrez]{moya2021exploring}
Borja Moya-Gómez, Marcin Stępniak, Juan~Carlos García-Palomares, Enrique Frías-Martínez, and Javier Gutiérrez.
\newblock Exploring night and day socio-spatial segregation based on mobile phone data: The case of medellin (colombia).
\newblock \emph{Computers, Environment and Urban Systems}, 89:\penalty0 101675, 2021.

\bibitem[Shin(2021)]{shin2021spatial}
Eun~Jin Shin.
\newblock Spatial segregation of chinese immigrants in seoul, south korea, during the covid-19 pandemic: Evidence from population data derived from mobile phone signals.
\newblock \emph{The Social Science Journal}, pages 1--22, 2021.

\bibitem[Silm et~al.(2021)Silm, Mooses, Puura, Masso, Tominga, and Saluveer]{silm2021relationship}
Siiri Silm, Veronika Mooses, Anniki Puura, Anu Masso, Ago Tominga, and Erki Saluveer.
\newblock The relationship between ethno-linguistic composition of social networks and activity space: A study using mobile phone data.
\newblock \emph{Social Inclusion}, 9\penalty0 (2):\penalty0 192--207, 2021.

\bibitem[Xian et~al.(2022)Xian, Qi, and Yip]{xian2022beyond}
Shi Xian, Zhixin Qi, and Ngai-ming Yip.
\newblock Beyond home neighborhood: Mobility, activity and temporal variation of socio-spatial segregation.
\newblock \emph{Journal of transport geography}, 99:\penalty0 103304, 2022.

\bibitem[Kronenberger and De~Saboya(2017)]{kronenberger2017configurational}
Bruna Da~Cunha Kronenberger and Renato~Tibiri{\\c{c}}{\'a} De~Saboya.
\newblock A configurational study of sociospatial segregation in the metropolitan region of florianopolis, brazil.
\newblock In \emph{Proceedings 11th International Space Syntax Symposium}, volume~11, pages 75--1, 2017.

\bibitem[Atuesta et~al.(2018)Atuesta, Ibarra-Olivo, Lozano-Gracia, and Deichmann]{atuesta2018access}
Laura~H Atuesta, J~Eduardo Ibarra-Olivo, Nancy Lozano-Gracia, and Uwe Deichmann.
\newblock Access to employment and property values in mexico.
\newblock \emph{Regional Science and Urban Economics}, 70:\penalty0 142--154, 2018.

\bibitem[{\"O}sth et~al.(2018){\"O}sth, Shuttleworth, and Niedomysl]{osth2018spatial}
John {\"O}sth, Ian Shuttleworth, and Thomas Niedomysl.
\newblock Spatial and temporal patterns of economic segregation in sweden’s metropolitan areas: A mobility approach.
\newblock \emph{Environment and Planning A: Economy and Space}, 50\penalty0 (4):\penalty0 809--825, 2018.

\bibitem[Schnell and Haj-Yahya(2014)]{schnell2014arab}
Izhak Schnell and Nasreen Haj-Yahya.
\newblock Arab integration in jewish-israeli social space: does commuting make a difference?
\newblock \emph{Urban Geography}, 35\penalty0 (7):\penalty0 1084--1104, 2014.

\bibitem[Zhou and Cheng(2019)]{zhou2019between}
Xiang Zhou and Yuning Cheng.
\newblock Between state and family: discussion on the segregation and integration of the daily living space within shanghai historic lane neighborhood.
\newblock \emph{Home Cultures}, 16\penalty0 (3):\penalty0 163--190, 2019.

\bibitem[Dorman et~al.(2020)Dorman, Svoray, and Kloog]{dorman2020does}
Michael Dorman, Tal Svoray, and Itai Kloog.
\newblock How does socio-economic and demographic dissimilarity determine physical and virtual segregation?
\newblock \emph{Journal of Spatial Information Science}, \penalty0 (21):\penalty0 177--202, 2020.

\bibitem[Collins(2004)]{collins2004interaction}
Randall Collins.
\newblock \emph{Interaction ritual chains}.
\newblock Princeton university press, 2004.

\bibitem[Pereira et~al.(2017)Pereira, Schwanen, and Banister]{pereira2017distributive}
Rafael~HM Pereira, Tim Schwanen, and David Banister.
\newblock Distributive justice and equity in transportation.
\newblock \emph{Transport reviews}, 37\penalty0 (2):\penalty0 170--191, 2017.

\bibitem[Aslak and Alessandretti(2020)]{aslak2020infostop}
Ulf Aslak and Laura Alessandretti.
\newblock Infostop: scalable stop-location detection in multi-user mobility data.
\newblock \emph{arXiv preprint arXiv:2003.14370}, 2020.

\bibitem[Hariharan and Toyama(2004)]{hariharan2004project}
Ramaswamy Hariharan and Kentaro Toyama.
\newblock Project lachesis: parsing and modeling location histories.
\newblock In \emph{International Conference on Geographic Information Science}, pages 106--124. Springer, 2004.

\bibitem[{Statistikmyndigheten SCB}(2023)]{SCB-deso}
{Statistikmyndigheten SCB}.
\newblock {DeSO – Demografiska statistikområden}, 2023.
\newblock URL \url{https://www.scb.se/hitta-statistik/regional-statistik-och-kartor/regionala-indelningar/deso---demografiska-statistikomraden/}.

\bibitem[Massey(2001)]{massey2001residential}
Douglas~S Massey.
\newblock Residential segregation and neighborhood conditions in us metropolitan areas.
\newblock \emph{America becoming: Racial trends and their consequences}, 1\penalty0 (1):\penalty0 391--434, 2001.

\bibitem[{Official Statistics of Sweden}(2016)]{travelsurvey}
{Official Statistics of Sweden}.
\newblock \emph{Swedish National Travel survey (RVU Sweden) 2011---2016}, 2016.
\newblock URL \url{https://www.trafa.se/en/travel-survey/travel-survey/}.

\bibitem[Seaman and White(2013)]{seaman2013review}
Shaun~R Seaman and Ian~R White.
\newblock Review of inverse probability weighting for dealing with missing data.
\newblock \emph{Statistical methods in medical research}, 22\penalty0 (3):\penalty0 278--295, 2013.

\bibitem[Van~de Kerckhove et~al.(2014)Van~de Kerckhove, Mohadjer, and Krenzke]{van2014weight}
Wendy Van~de Kerckhove, Leyla Mohadjer, and Thomas Krenzke.
\newblock A weight trimming approach to achieve a comparable increase to bias across countries in the programme for the international assessment of adult competencies.
\newblock \emph{JSM Proceedings, Survey Research Methods Section. Alexandria, VA: American Statistical Association}, pages 655--666, 2014.

\bibitem[{Google}(2019)]{GTFS_s}
{Google}.
\newblock {GTFS static dataset}, 2019.
\newblock URL \url{https://gtfs.org/reference/static}.

\bibitem[Wu et~al.(2021)Wu, Avner, Boisjoly, Braga, El-Geneidy, Huang, Kerzhner, Murphy, Niedzielski, Pereira, et~al.]{wu2021urban}
Hao Wu, Paolo Avner, Genevieve Boisjoly, Carlos~KV Braga, Ahmed El-Geneidy, Jie Huang, Tamara Kerzhner, Brendan Murphy, Micha{\l}~A Niedzielski, Rafael~HM Pereira, et~al.
\newblock Urban access across the globe: an international comparison of different transport modes.
\newblock \emph{npj Urban Sustainability}, 1\penalty0 (1):\penalty0 16, 2021.

\bibitem[Pereira et~al.(2021)Pereira, Saraiva, Herszenhut, Braga, and Conway]{pereira2021r5r}
Rafael~HM Pereira, Marcus Saraiva, Daniel Herszenhut, Carlos Kaue~Vieira Braga, and Matthew~Wigginton Conway.
\newblock r5r: rapid realistic routing on multimodal transport networks with r 5 in r.
\newblock \emph{Findings}, 2021.

\end{thebibliography}

\section*{Acknowledgements}
This research is funded by the Swedish Research Council (Project Number 2022-06215).

\section*{Author contributions}
Y.L. and L.A. conceptualized the study. 
Y.L., J.G., and R.H.M.P. processed the data. 
Y.L. conducted simulations and data analysis. 
All authors wrote the manuscript.

\section*{Funding}
Open access funding is provided by Chalmers University of Technology.

\section*{Competing interests}
The authors declare that there are no conflicts of interest.

\section*{Additional information}
Correspondence and requests for materials should be addressed to Y.L.

\appendix
\renewcommand{\thefigure}{S.\arabic{figure}}
\renewcommand{\thetable}{S.\arabic{table}}
\setcounter{figure}{0}
\setcounter{table}{0}
\renewcommand{\thefigure}{S.\arabic{figure}}
\renewcommand{\thetable}{S.\arabic{table}}
\setcounter{figure}{0}
\setcounter{table}{0}
\section{Data and processing}
\subsection{Mobility data}\label{seca:mobi_data}
Mobile application data consists of GPS records collected through location-enabled applications installed on people's smartphones, with the form of (\textit{id, lat, lon, time}). 
This study uses a dataset from a diverse set of mobile apps used by anonymized adult smartphone users in Sweden\footnote{http://www.pickwell.co/}. \par

\begin{figure}
\centering
\includegraphics[width=0.7\textwidth]{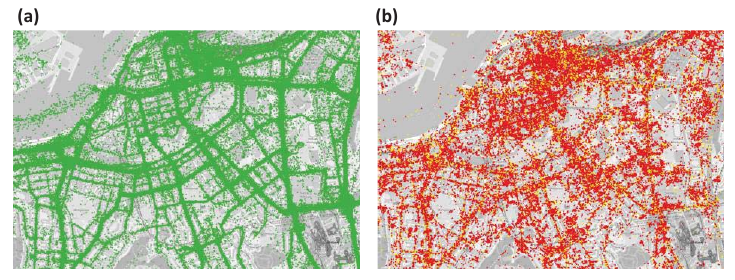}
  \caption{Geolocations from mobile application data in the central Gothenburg. \textbf{(a)} Movement points. \textbf{(b)} Stationary points.}\label{fig:raw_map}
\end{figure}

The dataset contains geolocations of moving (Figure \ref{fig:raw_map}a) and being stationary (Figure \ref{fig:raw_map}b) collected from cell towers, GPS, and Wi-Fi sources, depending on the geolocation source at the moment when specific phone applications being used. 
Therefore, this dataset features varying sampling frequencies and spatial resolutions. 
It also contains sampling bias, given that geolocations are collected only when the users use certain applications on their phones. \par

\subsection{Stay detection}\label{seca:stay_detection}
The Infostop algorithm has four configurable parameters to identify stays, summarized in Table \ref{tab:info_stop_paras}. 
The two spatial parameters, $r_1$ and $r_2$, control the spatial accuracy of identified stays; the greater their values, the less accurate the identified stays. 
Too small values of these two parameters make the algorithm sensitive to GPS noise. 
To balance spatial accuracy and robustness against noise, we choose 30 m based on the raw data's distribution. \par

The minimum duration of a stay is 15 min because we want to approximate the social interactions by identifying individuals' stays for activities.
A too-short stay is unlikely to trigger social interactions with those in the same place.
While an overly long minimum duration of a stay may miss some activities, especially considering the data's temporal sparsity issue. \par

The last parameter, $t_{max}$, affects the maximum stay duration and how many records will be included in one stay.
The raw geolocation data are passively collected only when people use mobile phone applications. 
Therefore, these geolocations are sparse in the time dimension and need careful exploration to determine a reasonable value for $t_{max}$ to avoid creating artifacts in the detected stays.
The artifacts originate from unobserved locations in the data between two consecutive geolocations close to each other in space.
If we pick a big $t_{max}$ to detect stays, these two close geolocations, interrupted by other movements in between but unobserved from the data, will be falsely merged into one stay. \par

\begin{figure}
\centering
\includegraphics[width=0.5\textwidth]{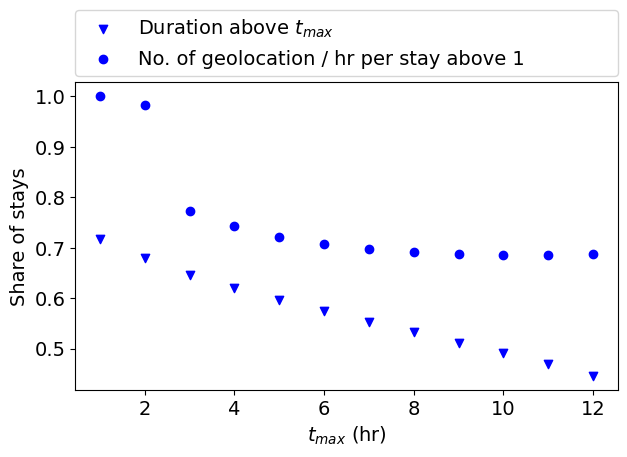}
  \caption{Impact of $t_{max}$ on the detected stays (a sample dataset).}\label{fig:t_max_infostop}
\end{figure}

To determine the $t_{max}$ value for stay detection, we test 1 to 12 hours in 1 hour intervals and quantify its impact on the detected stays.
In the lack of ground truth of individuals' whereabouts, we examine the resulting stays regarding their duration distribution, i.e., the share of stays with duration above $t_{max}$, and their underlying records, i.e., the percentage of stays with average geolocations per hour above 1.
Figure \ref{fig:t_max_infostop} shows the impact of $t_{max}$ on the detected stays. \par

We choose 3 hours for $t_{max}$ to detect stays. 
This ensures sufficient geolocations underlying a stay (c.a. 80\% stays above one geolocation per hour) and reasonable stay durations (35\% stays below 3 hours). 
According to the Swedish National Travel Survey, 37\% of activities are below 3 hours \cite{travelsurvey}. \par

\subsubsection{Data filtering}\label{seca:filtering}
The individuals meeting the below criteria are selected for further analysis.
\begin{itemize}
\item The identified home has a corresponding grid in Sweden's population grids and Demographic statistics areas (DeSO) zones. In other words, the individuals are presumed to live in Sweden.
\item One should have at least three nights at home.
\item The identified stays above 12 hours are removed.
\item The individuals should have more than seven active days, and the number of unique locations should be more than two.
\end{itemize}

\subsubsection{Data description}\label{seca:mobi_desc}
After processing the mobility data, we have 30,454,903 stays from 322,919 individual devices for further analysis.
Their stays have the characteristics shown in Table \ref{tab:stays_stats}. \par

\subsection{Census data}\label{seca:census_data}
DeSO - Demographic statistical areas are a national subdivision defined by Statistics Sweden, following Sweden's county and municipal boundaries \cite{SCB-deso}. 
There are 5,984 DeSO zones, each with a unique 9-digit code, and the first three digits indicate which county it belongs to. 
The fifth position is a letter: A, B, or C, which groups DeSO zones into three different categories, where in this study, A and B stand for Rural/Suburban and C for Urban. \par

DeSO zones have varying sizes; however, these zones are designed to reflect the population distribution.
Urban areas with high population densities have a high spatial resolution, i.e., they are represented by many small DeSO zones.
Therefore, this study uses DeSO zones as the spatial unit to calculate visiting and experienced segregation metrics. \par

The statistics agency in Sweden (SCB) also provides grid-level information based on the total population register. 
We collect these population grids from the Swedish University of Agricultural Sciences\footnote{\url{https://maps.slu.se/}}.
To protect privacy, there are two spatial resolutions: 250 m x 250 m (small) and 1000 m x 1000 m (large). 
The used population grids have 178,920 cells, where 57\% are large, and 43\% are small grid cells.
The population size in each grid ranges between 0 and 7,844, with a median value of 13. 
We keep these 7\% zero-population grid cells because they have non-zero daytime jobs used for transport access computation. \par

\subsection{Analysis zones}\label{seca:ana_zones}
To balance data sufficiency and analysis accuracy, the analysis zones for this study combine the census zones and the H3 zoning system\footnote{Hexagonal hierarchical geospatial indexing system. \url{https://h3geo.org/}}.
Within each census zone, we create its children hexagons under H3, a geospatial indexing system that partitions the world into hexagonal cells.
The size/resolution of the hexagons within each census zone depends on the census zone's area, as detailed in Table \ref{tab:seca_ana_zones}.

\subsection{Sampling debiasing}\label{seca:samp_debias}
\begin{figure}
\centering
\includegraphics[width=0.5\textwidth]{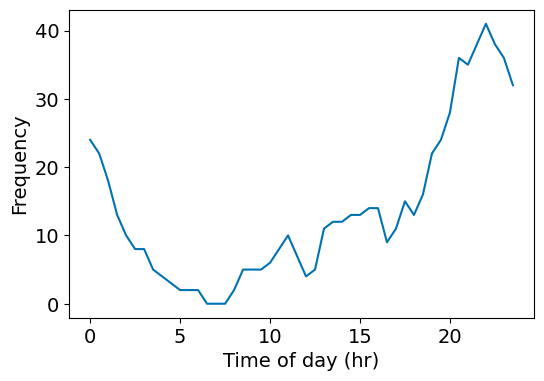}
  \caption{Temporal distribution of detected stays of an exemplary individual.}\label{fig:stays_tempo_indi}
\end{figure}

Unlike travel surveys or constant GPS tracking, mobile phone application data only gives the population's geolocation when they use certain applications.
This passive data collection results in imbalanced knowledge of where the individuals are (Figure \ref{fig:stays_tempo_indi}).
To reduce such bias, this step assigns a weight to each detected stay individually, considering the overall temporal patterns of each device. \par

For a given individual device, we first divide 24 hours into 48 groups with evenly placed half-hour intervals and then calculate the number of observed locations in each interval ($f_i$, i=1,2,...,48).
Next, each time interval is assigned a weight ($w_i=1/f_i$, i=1,2,...,48).
Finally, we calculate how many intervals a detected stay spans over, e.g., a stay starting from 8 am till 9 am spans over the groups ${17,18}$.
And its weight is defined as $W=w_{17}+w_{18}$.
The stay weights mimic the even sampling of a person's whereabouts, leading to a statistically smooth timeline. \par

\subsection{Population debiasing}\label{seca:pop_debias}

After identifying the individuals' homes, we know how many live in each DeSO zone.
The Spearman correlation between the number of devices and the population size has a coefficient of 0.47 ($p<0.001$), indicating mobile phone data's magnitude-wise representation of the actual population. \par

\begin{figure}
\centering
\includegraphics[width=0.8\textwidth]{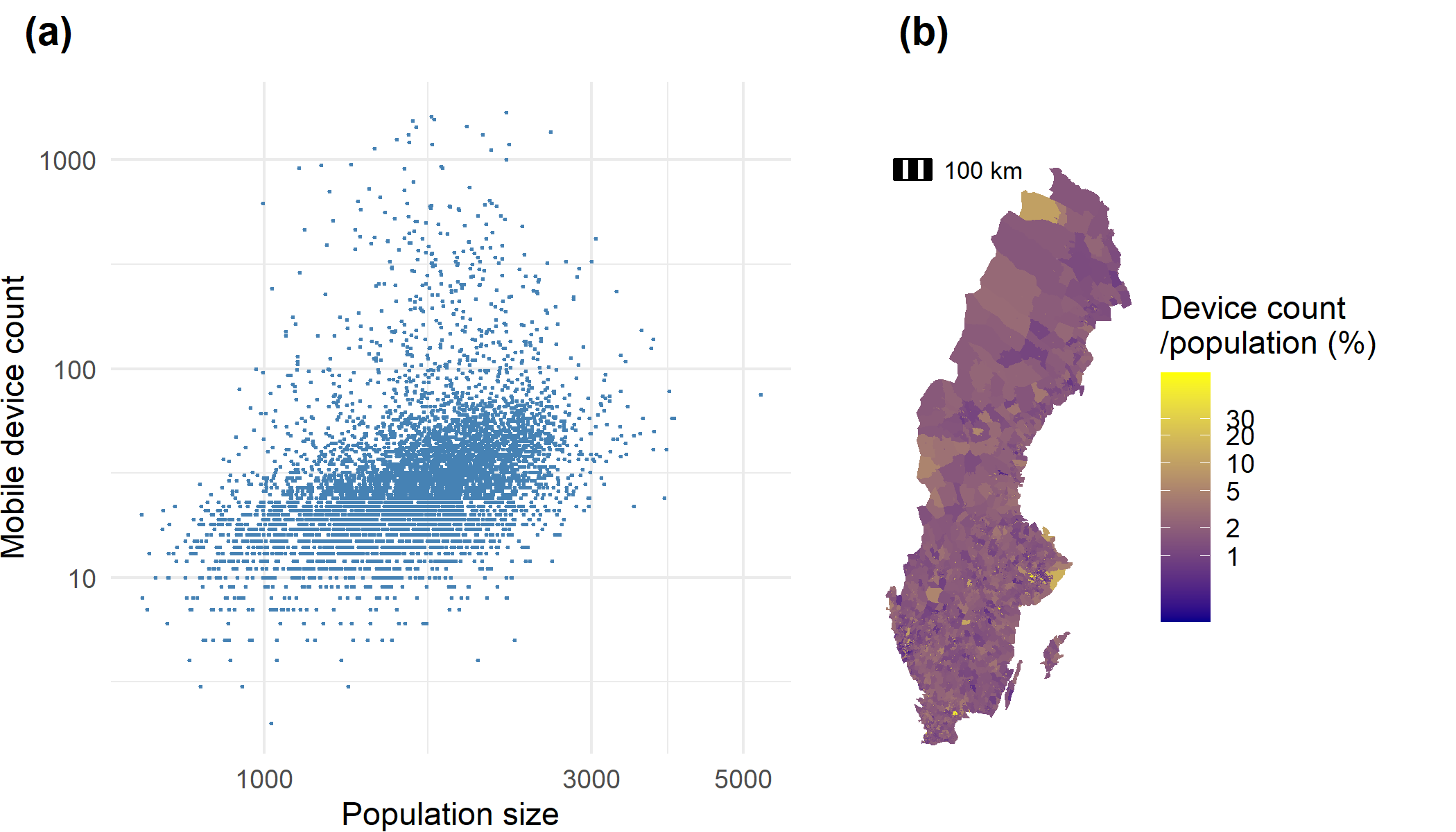}
  \caption{Home distribution. \textbf{(a)} Number of devices vs. actual population size by DeSO zone. \textbf{(b)} Share of devices compared to the actual population by DeSO zone.}\label{fig:homes_desc}
\end{figure}

Figure \ref{fig:homes_desc} shows the distributions of individual device numbers compared to the actual population sizes. 
Figure \ref{fig:homes_desc}a confirms the medium-level correlation between the individual device number and the population size in DeSO zones.
Figure \ref{fig:homes_desc}b shows to what extent DeSO zones' populations are represented in the mobile phone data.
The share of the population represented ranges between 0.19\% and 93\%, except for six outlier zones with more devices than the actual population. \par

We use inverse probability weighting (IPW) to assign each device a weight \citep{seaman2013review}, i.e., the reverse of the phone users' count ratio to the DeSO zone's actual population size ($W_{p}$).
This individual weight, $W_{p}$, has extreme values in some DeSO zones where only a few devices are included. 
Weight trimming technique is applied \citep{van2014weight}, where any weight above the cut-point weight ($W_0$) is set to $W_0$. 
Equation \ref{eq:weight_0} determines this cut-point weight value. \par

\begin{equation}
W_0=3.5\sqrt{1+{\tt CV}^2\left({\bf W_p}\right)}\times {\tt Med} \left({\bf W_p}\right)\label{eq:weight_0}
\end{equation}%
where $\tt CV$ is the coefficient of variance and $\tt Med$ is the median value.

\subsection{Transportation data}
Points of interest and road networks are retrieved from OpenStreetMap archives of Sweden\footnote{https://download.geofabrik.de/europe.html}.

\subsubsection{Point of interest data}\label{seca:pois}

Each point of interest (POI) has a class and subclass defined by OSM contributors.
For example, a geolocation is labeled as ``class=amenity'' and ``subclass=toilets.''
There are over a thousand subclasses in the original Swedish POI dataset.
For the ease of simulation, this study cleans the POI dataset and groups them into broader but smaller number of categories. \par

We first use GPT-4 to learn 24 preliminary categories based on the POIs' subclass labels of Sweden.
Then we manually check the validity of this preliminary category assigned to each POI's subclass.
We also exclude those POIs such as emergency point, toilet, etc. where people spend limited time and the social interactions are unlikely to happen.
In addition, we merge some classes: tourism, historic = tourism, leisure, sport = leisure, craft, office = office, and distinguish POIs of the same sublcass but different classes of amenity or shop.
At last, there are 33 categories of POIs where a stands for amenity and s stands for shop: Artisan Workshops, Automotive Services (a), Automotive Services (s), Craft, Education (a), Education (s), Entertainment (s), Fashion and Accessories (s), Financial Services (a), Financial Services (s), Food and Drink (a), Food and Drink (s), Groceries and Food (a), Groceries and Food (s), Health and Beauty (a), Health and Beauty (s), Healthcare (a), Healthcare (s), Home and Living, Leisure, Office, Office (s), Outdoor Recreation (a), Outdoor Recreation (s), Recreation (a), Recreation (s), Religious Places (a), Shop, Sports and Activities (a), Sports and Activities (s), Tourism, Transportation (a), Transportation (s). \par

Randomly shifting original stay records to a nearby similar kind of POI follows a few steps.
The candidate nearby POIs shall belong to the same category. 
If none is found, then we do not distinguish amenity and shop categories, Office and Craft, or any categories belonging to any shops and the overall Shop category. 
The shifted POI is randomly selected from the POI candidate set, if any.

\subsubsection{Public transit data}\label{seca:gtfs}
General Transit Feed Specification (GTFS) is an open data standard for public transport timetables proposed by Google. 
A GTFS static dataset \citep{GTFS_s} is a collection of text files consisting of all the information required to reproduce a transit agency's schedule, including the locations of stops and timing of all routes and vehicle trips. 
This project collects the up-to-date GTFS data of Sweden from Samtrafiken AB\footnote{https://samtrafiken.se/}.
The GTFS data are processed to remove the invalid shapes, extract the geolocations of all the transit stops, and calculate access to jobs by public transport. \par

\subsection{Transport access}\label{seca:trans_ind}
Transport access quantifies how easily one can reach various destinations. 
In this study, we consider taking public transit from the residence grid. 
We use a cumulative-opportunity indicator \citep{wu2021urban} to measure the number of jobs accessible within 30 min travel by public transport ($A_t$).
The other relevant parameters, such as walking speed, are the default settings of \textit{accessibility} function in \textit{r5r}, an open-source transportation analysis package \citep{pereira2021r5r}. \par

\section{Modeling segregation by birth background}
\subsection*{Adjusted ICE vs. its original form}\label{secb:ice_comp}
The original formula of ICE is shown below (Equation \ref{eq:ice_o}).
\begin{equation}
ICE_j = \frac{N_j - F_j}{N_j + F_j + O_j} \label{eq:ice_o}
\end{equation}

\begin{figure}
\centering
\includegraphics[width=0.5\textwidth]{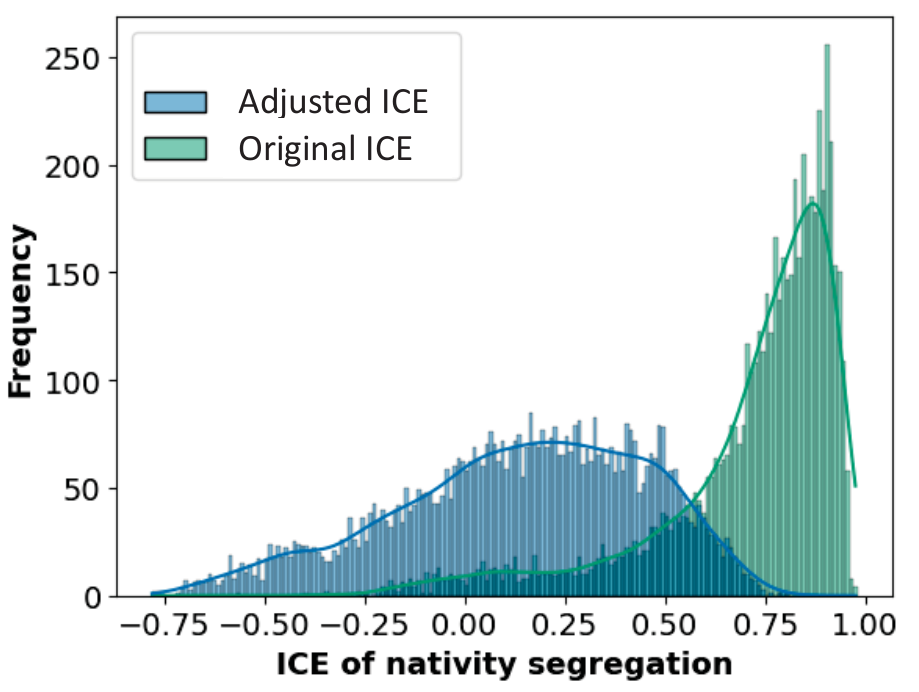}
  \caption{Residential segregation by birth background: adjusted vs. original.}\label{fig:ice_metrics}
\end{figure}

For the comparison, Figure \ref{fig:ice_metrics} shows the distribution of DeSO zones regarding the two ICE metrics using Equations \ref{eq:ice_o} in comparison with the one used in this study (Equation 1 in the research report). \par

\subsection{Individual attributes}\label{secb:indi_attr}
\begin{figure}
\centering
\includegraphics[width=1\textwidth]{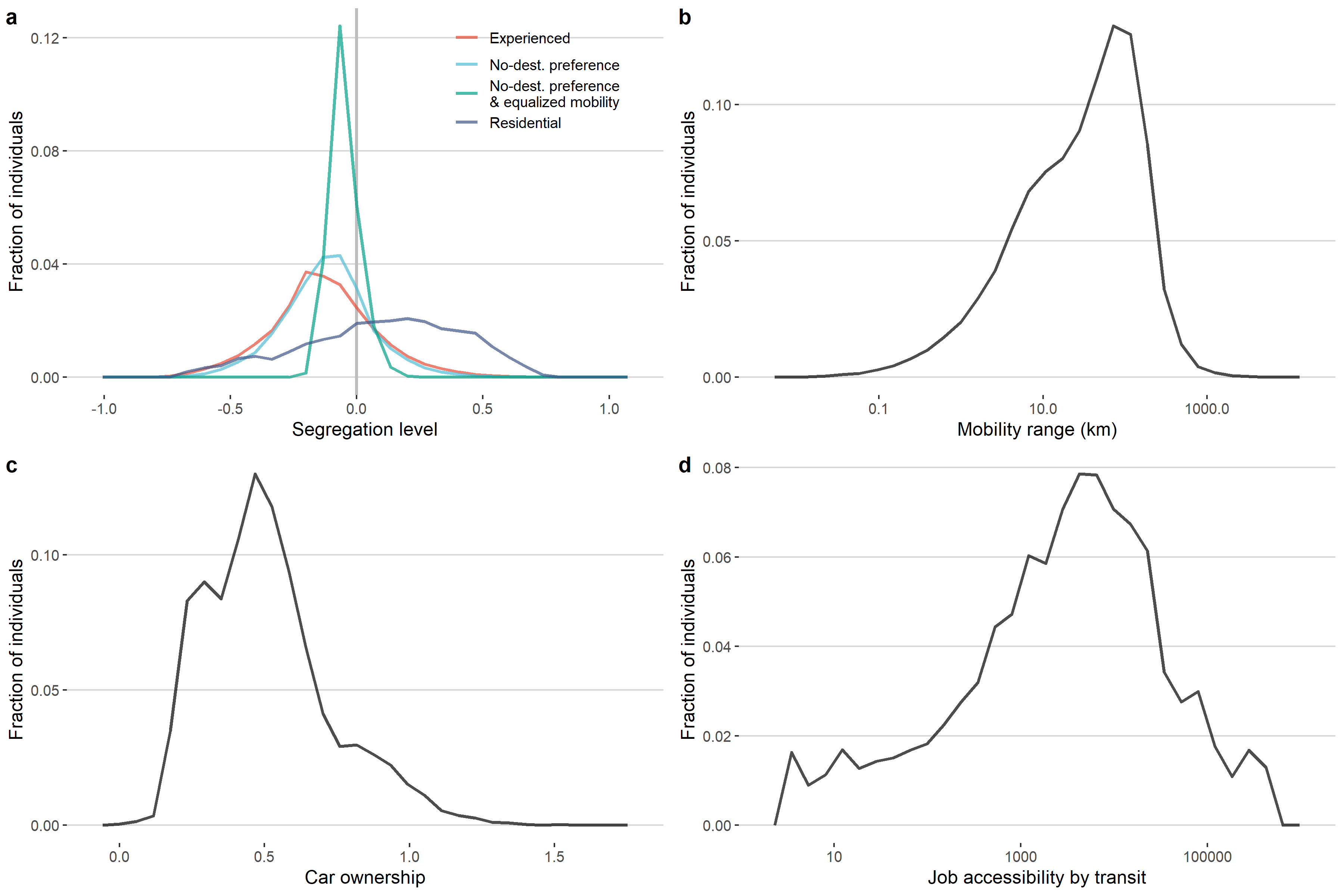}
  \caption{Weighted individual distribution of key attributes. \textbf{a}, Experienced, residential, and simulated segregation levels. \textbf{b}, Mobility range. \textbf{c}, Car ownership per capita. \textbf{d}, Job accessibility by transit.}\label{fig:seg_trans_desc}
\end{figure}

Figure \ref{fig:seg_trans_desc} shows the distributions of individual attributes.
Compared with residential segregation, the experienced and simulated segregation decrease collectively. 
Car ownership is predominantly skewed toward smaller values with a long tail.
The number of accessible jobs by transit is generally smaller than that by car. 
At the same time, a few regions located at the very center of the cities have an exceptionally high density of transit services. \par

\subsection{Residential vs. experienced segregation}\label{secb:seg_comp}
\begin{figure}
\centering
\includegraphics[width=0.6\textwidth]{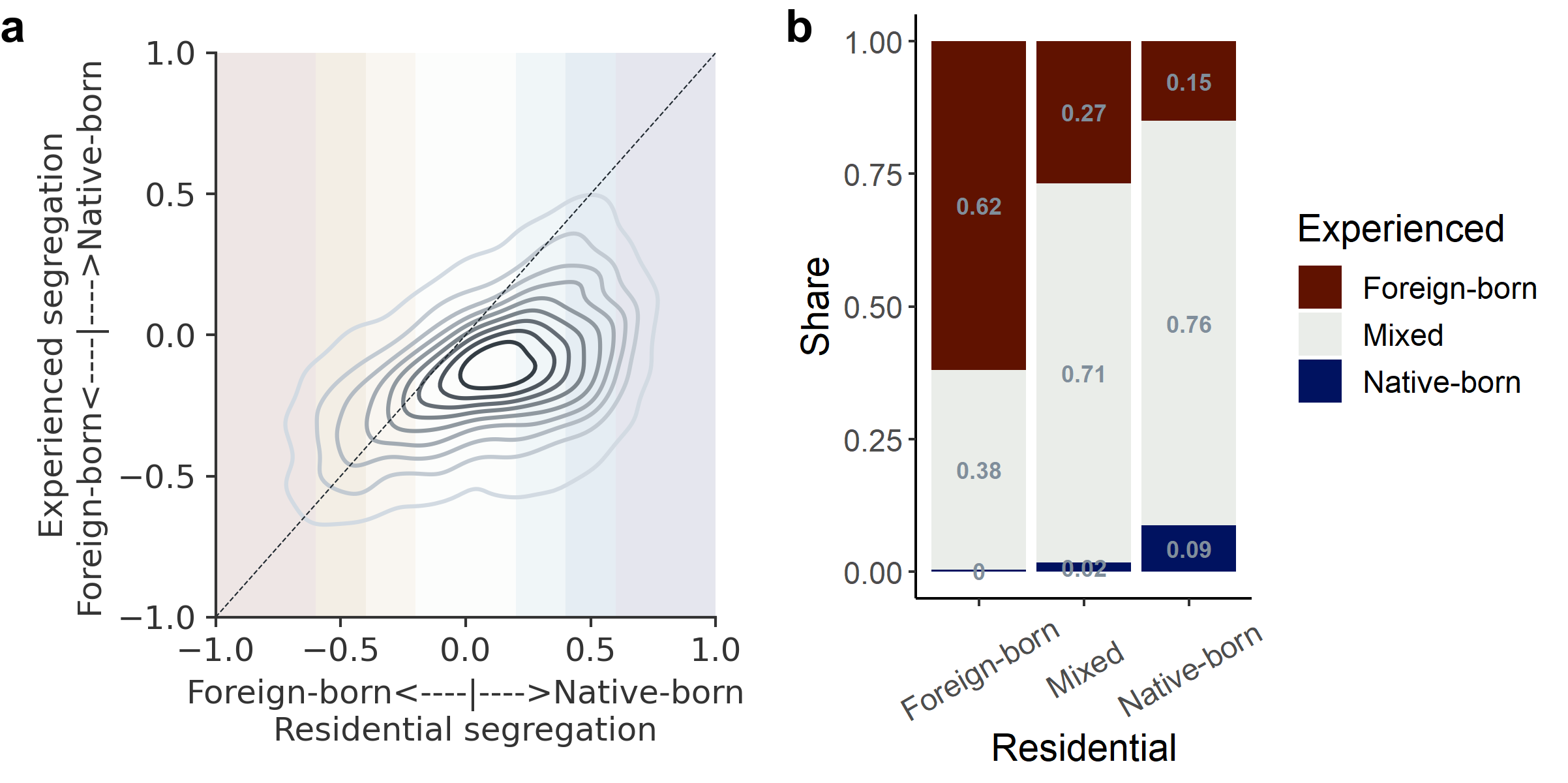}
  \caption{Residential segregation and experienced segregation outside residential area. \textbf{a}, Scatter density plot of Experienced ($ICE_e$) vs Residential ($ICE_r$) segregation for all individuals in the dataset. Background colors indicate levels of residential segregation (see color bar in subplot a). The gray contour lines represent areas with similar data point densities, as determined by a kernel density estimate (KDE) plot. The diagonal line indicates where $ICE_e=ICE_r$. \textbf{b}, Share of individuals by experienced segregation ($ICE_e$, y-axis) across different residential segregation groups ($ICE_r$, x-axis). Individuals are categorized by residential segregation into three groups: $ICE_r < -0.2$ (foreign-born), $ICE_r > 0.2$ (native-born), and $-0.2 \leq ICE_r \leq 0.2$ (mixed). The same categorization criteria are applied to their experienced segregation along the y-axis. }\label{fig:res_vs_exp}
\end{figure}

To assess the impact of mobility on individual segregation ($ICE_e$ vs. $ICE_r$), we categorize individuals into three groups.  
This classification mirrors the previously employed residential segregation methodology. 
Specifically, individuals are grouped based on their $ICE_e$ as follows: those with $ICE_e>0.2$ are categorised as native-born segregated, those with $ICE_e<-0.2$ as foreign-born segregated, and those within the range $-0.2 \leq ICE_e \leq 0.2$ as exposed to a mixed group of individuals (see Fig. \ref{fig:res_vs_exp}b). 
Our analysis reveals notable shifts between residential and experienced segregation (see Fig. \ref{fig:res_vs_exp}b).
Among individuals initially identified as native-born (Group N) based on residential data, $76\%$ transition into the mixed category (Group M) when mobility is considered. 
Conversely, only $38\%$ of individuals initially identified as foreign-born (Group F) transition to the mixed category. 
The remaining $62\%$ of individuals in this group continue to experience a strong and significant level of foreign-born segregation during their daily activities (see Table \ref{tab:ice_r_stats}). 

\subsection{Impact of mobility range}\label{secb:mobi_cap}
Individuals in foreign-born segregated areas exhibit distinct characteristics regarding car ownership, transport access, and their experienced segregation, in contrast to those living in native-born concentrated areas (Table \ref{tab:access_stats}).
Over 50\% of individuals in Group N have a high car ownership, while in Group F, the majority have a low to medium car ownership.
Car ownership negatively correlates with job accessibility by transit.
Group F has lower public transport access than Group N, except for those with high car ownership.
This is especially drastic for those with low levels of car ownership. 
These individuals living in native-born segregated areas have almost nine times more accessible jobs by taking transit than their foreign-born counterparts.
This may be related to the distinction between Groups N and F regarding where they live, primary travel mode, economic circumstances, and individual preferences. \par

\subsection{Statistical test results}\label{secb:stats}
\begin{figure}
\centering
\includegraphics[width=1\textwidth]{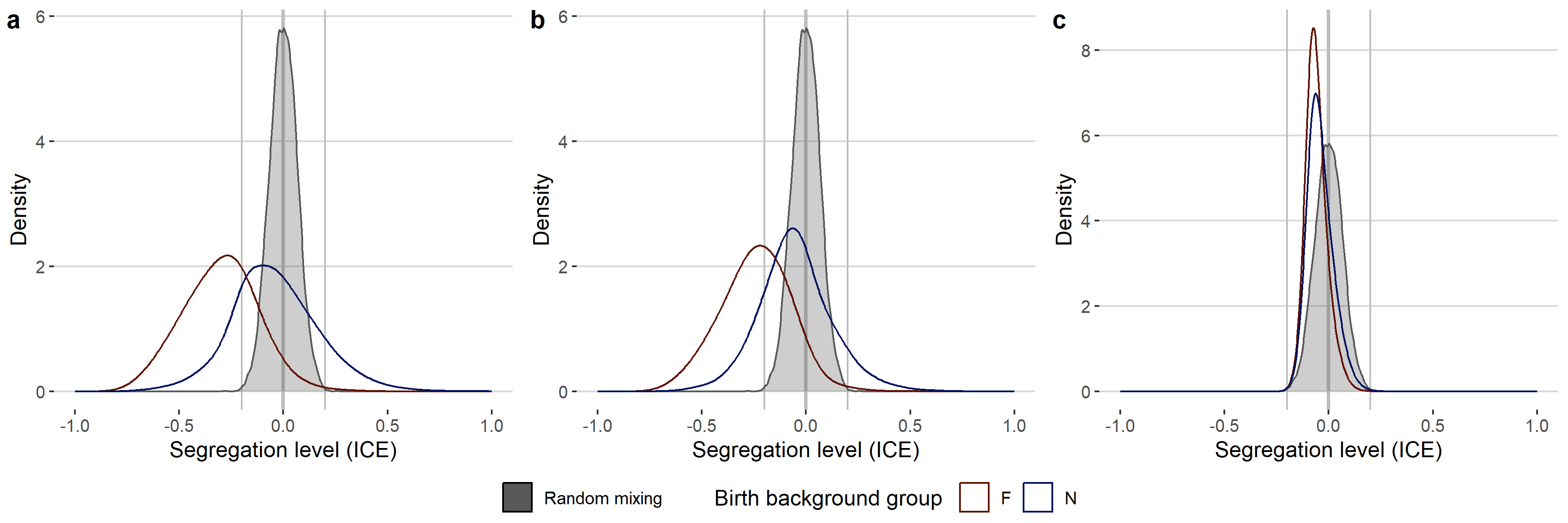}
  \caption{Experienced segregation levels outside residential area: comparison with Random mixing. Vertical lines mark the thresholds, -0.2 and 0.2, for the statistical test to determine whether the distribution is significantly segregated. \textbf{a}, Experienced. \textbf{b}, No dest. Preference. \textbf{c}, Equalized mobility \& no dest. preference.}\label{fig:tests_ice}
\end{figure}

\begin{figure}
\centering
\includegraphics[width=1\textwidth]{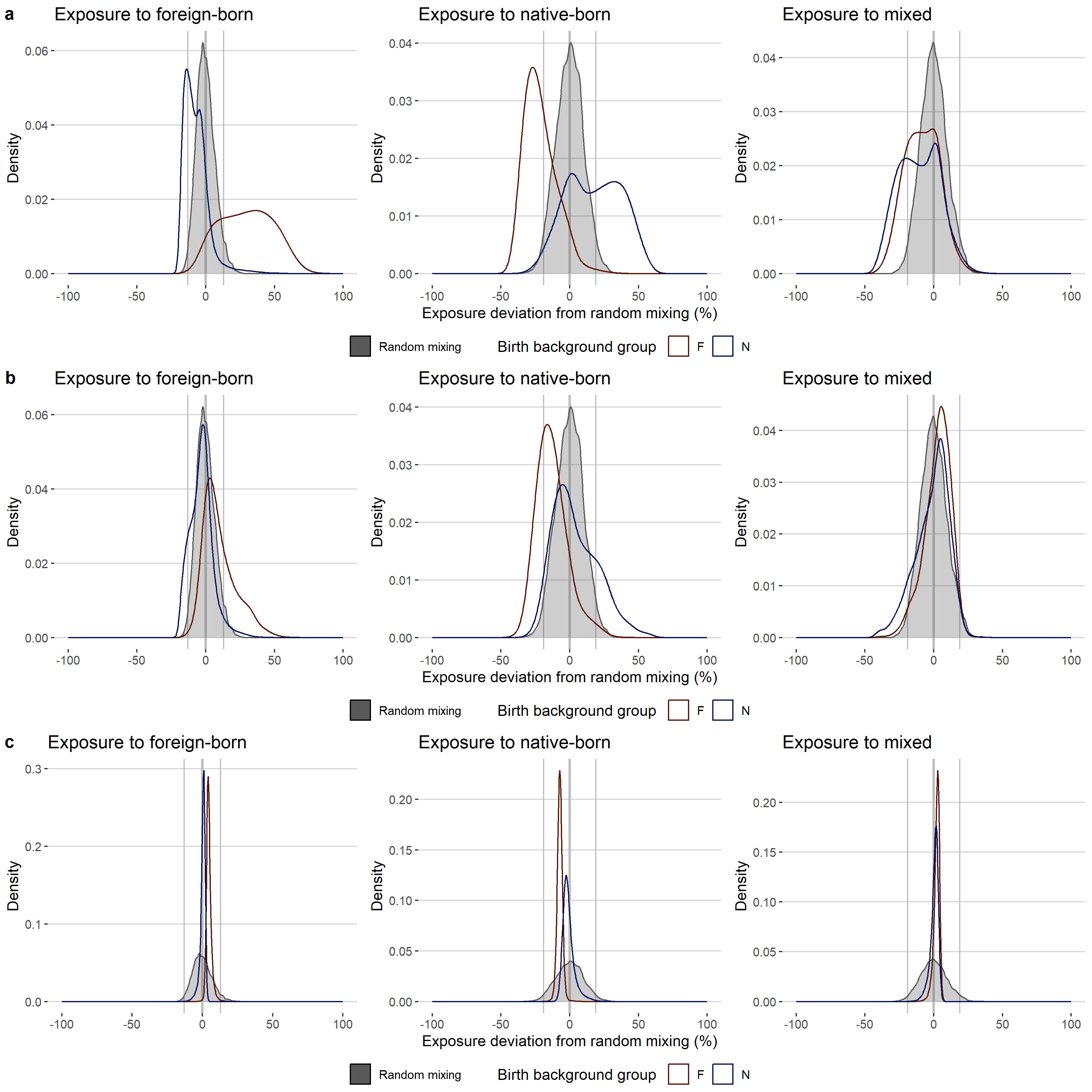}
  \caption{Group exposure outside residential area: comparison with Random mixing. Vertical lines mark the thresholds, -13 and 13 for F, and -19 and 19 for N and M, for the statistical test to determine whether the distribution deviates significantly from Random mixing. \textbf{a}, Experienced. \textbf{b}, No dest. Preference. \textbf{c}, Equalized mobility \& no dest. preference.}\label{fig:tests_exp}
\end{figure}

We present a summary of segregation metrics and group exposure statistics, including statistical test results. 
All median values are calculated using a weighted bootstrap method with 1000 repetitions, and the error is represented by the standard deviation of the bootstrap estimates. 
The statistical tests are conducted using the Wilcoxon weighted test to determine significant deviations from the baselines: segregation (Tables \ref{tab:ice_r_stats}-\ref{tab:ice_e_stats} and Figure \ref{fig:tests_ice}) and deviation from random mixing group exposure (Table \ref{tab:inter_stats} and Figure \ref{fig:tests_exp}). \par

\begin{figure}
\centering
\includegraphics[width=0.6\textwidth]{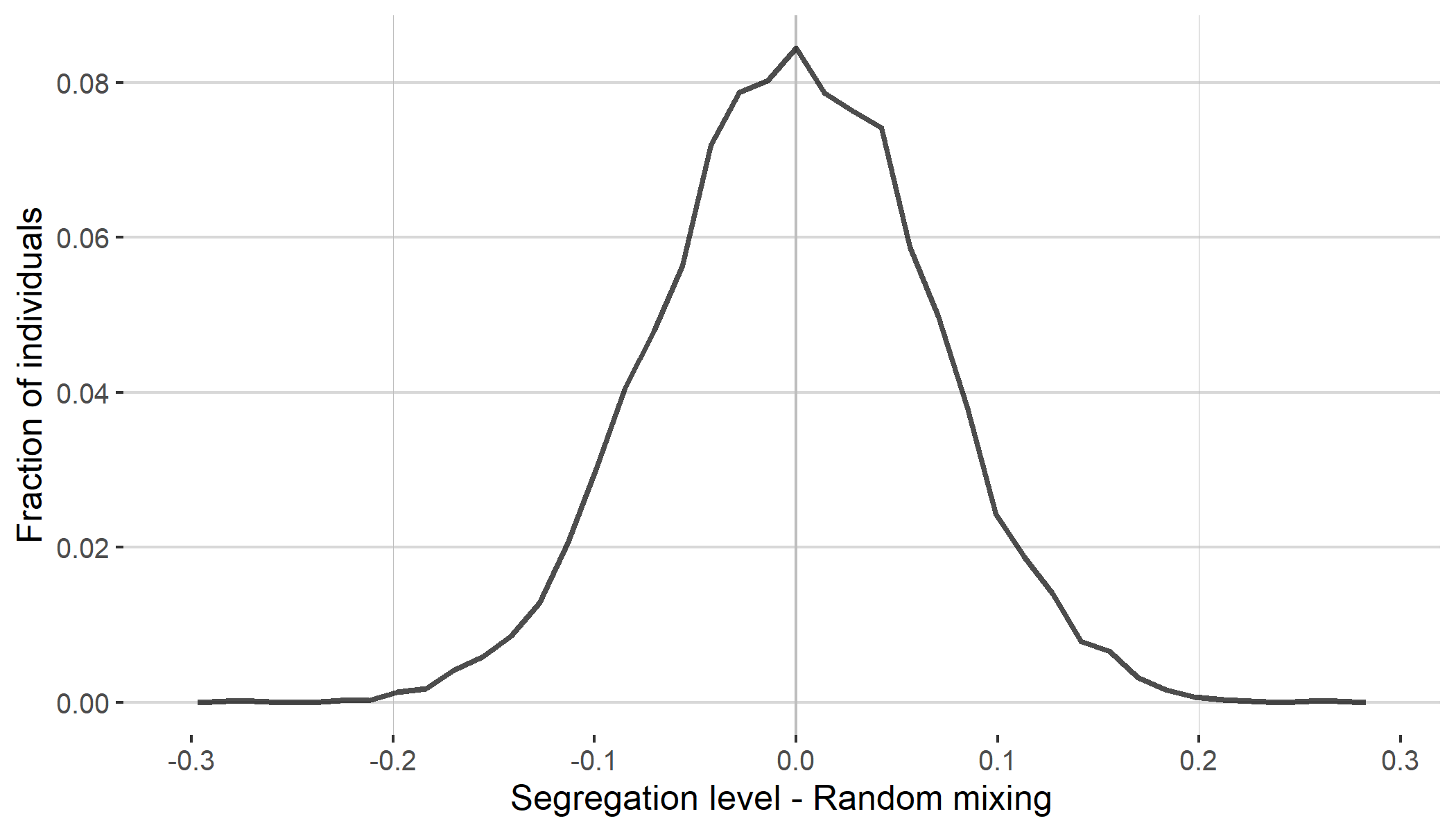}
  \caption{Distribution of residential segregation levels of individuals in a simulated scenario where individuals randomly choose where to live in Sweden. The distribution contains 100 repetitions of the simulation. The data between the two vertical lines correspond to the 99\% confidence interval around 0.}\label{fig:seg_rsim}
\end{figure}

\begin{figure}
\centering
\includegraphics[width=0.6\textwidth]{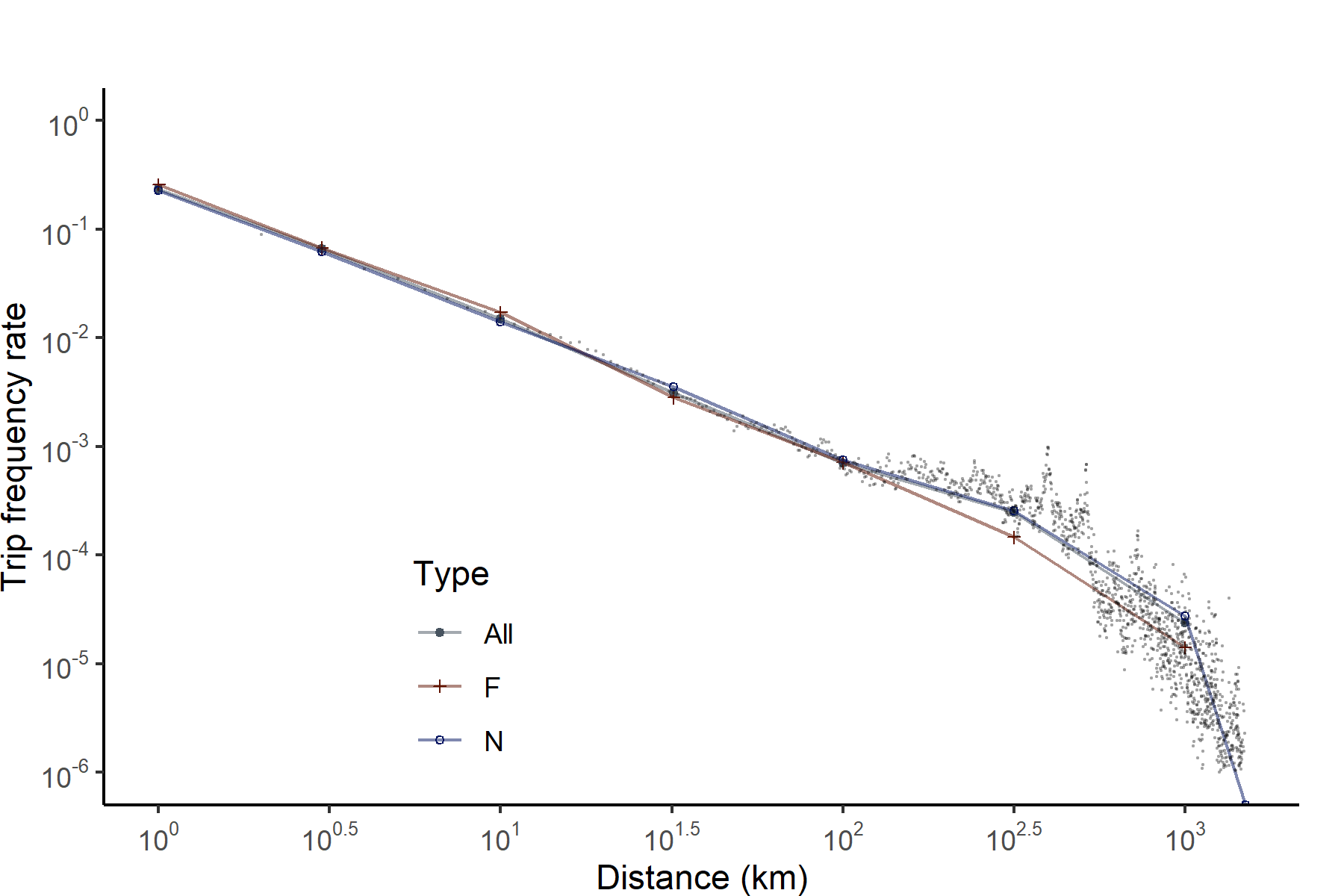}
  \caption{Distribution of distance-to-home of visited places. The gray points represent the frequency rates of all individuals across 1,500 distance groups. The larger points and lines depict results from selected distance groups, accentuating data trends and contrasting the differences between foreign-born and native-born.}\label{fig:limited_travel_dist}
\end{figure}

\begin{figure}
\centering
\includegraphics[width=1\textwidth]{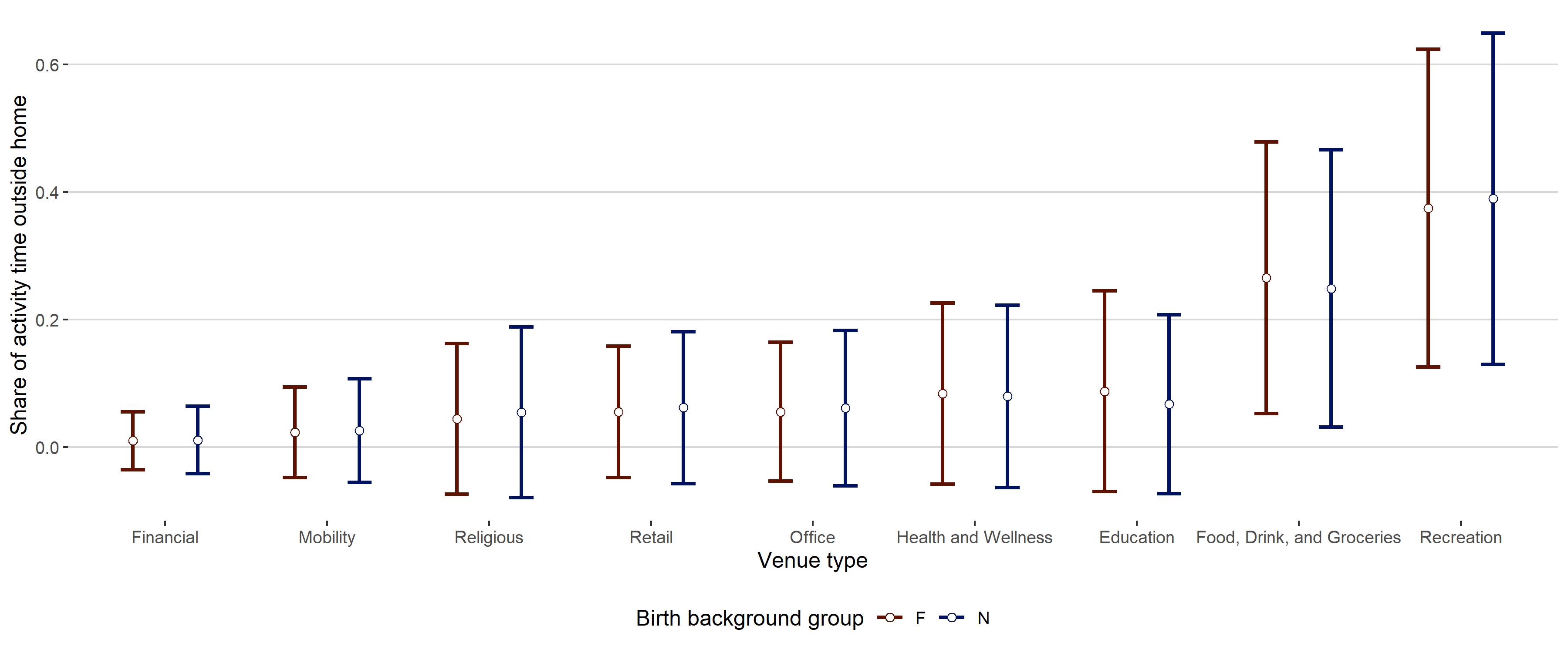}
  \caption{Activity patterns: share of activity time by venue type and birth background group. Lines show the ranges between the values of weighted mean $\pm$ standard deviation.}\label{fig:activity}
\end{figure}

\begin{table}\centering
\caption{Infostop algorithm parameters and set values.}\label{tab:info_stop_paras}
\begin{tabularx}{\textwidth}{lXll}
Parameter & Definition                                                                                                  & Unit  & Value \\\midrule
$r_1$ & The maximum roaming distance allowed for two points within the same stay                                   & meter & 30    \\
$r_2$ & The typical distance between two stays at the same destination                                             & meter & 30    \\
$t_{min}$ & The minimum duration of a stay                                                                              & min   & 15    \\
$t_{max}$ & The maximum time difference between two consecutive records to be considered within the same stay & hour  & 3\\
\bottomrule
\end{tabularx}
\end{table}

\begin{table}\centering
\caption{Descriptive stay statistics. An active day is when at least one stay is detected.}\label{tab:stays_stats}
\begin{tabular}{lrrrrr}
Attribute                  & min & 25\% & 50\% & 75\% & max  \\\midrule
No. of unique locations    & 3   & 4    & 7    & 18   & 594  \\
No. of active days        & 8   & 21   & 37   & 68   & 215  \\
No. of stays               & 5   & 27   & 54   & 117  & 1882 \\
Median stay duration (min) & 20  & 180  & 183  & 194  & 4207\\
\bottomrule
\end{tabular}
\end{table}

\begin{table}\centering
\caption{Analysis zones created combining parental census zones and children hexagons. There are 70,830 analysis zones, contrasting with the 5,984 census zones.}\label{tab:seca_ana_zones}
\begin{tabularx}{\textwidth}{lllX}
Parental census zone area (km$^2$) & No. of census zones & Children H3 zone resolution & Average area of analysis zones (km$^2$) \\\midrule
\textless 0.5                   & 1509                & None$^1$                        & 0.2                            \\
0.5-3.5                         & 1817                & 9                           & 0.1                            \\
3.5-15                          & 1424                & 8                           & 0.7                            \\
15-100                          & 440                 & 7                           & 5.2                            \\
100-720                         & 683                 & 6                           & 36.1                           \\
\textgreater 720                & 111                 & 5                           & 252.9  \\
\bottomrule
\end{tabularx}
\footnotetext[1]{The original census zones below 0.5 km$^2$ are preserved in the analysis zones.}
\end{table}

\begin{table}\centering
\caption{Residents of the foreign-born and native-born segregated areas: average values of key statistics in segregation and transport access considering car ownership levels. Car ownership is divided into three groups representing the lowest 25\%, the middle 50\%, and the highest 25\%: Low ($<$ 0.28), Medium (0.28--0.72), and High ($>$0.72). F = Individuals living in foreign-born segregated areas, N = Individuals living in native-born segregated areas.}\label{tab:access_stats}
\begin{tabular}{llrrr}
Group              & Car ownership & Share in Group (\%) & Job access by transit ($\times 10^3$) & Experienced segregation \\\midrule
\multirow{3}{*}{F} & Low           & 38.4          & 17.8                  & -0.32 \\
                   & Medium        & 58.8           & 6.2                   & -0.29 \\
                   & High          & 2.8            & 1.3                   & -0.21 \\\hline
\multirow{3}{*}{N} & Low           & 2.1           & 203.4                & -0.05  \\
                   & Medium        & 39.3           & 3.2                   & -0.06  \\
                   & High          & 58.6           & 0.1                   & -0.05 \\
\bottomrule
\end{tabular}
\end{table}

\begin{table}\centering
\caption{Residential segregation level ($ICE_r$) by group. Errors are computed as the median values' bootstrap standard deviation (1000 repetitions). $^*$ This is defined by 99\% confidence interval around zero in the randomized mixing scenario. See details in Section Identifying Segregated Individuals.}\label{tab:ice_r_stats}
\begin{tabular}{lrrl}
Group & Median & Error & Segregation$^*$ \\\midrule
F     & 0.385  & 0.001 & Yes         \\
N     & -0.380 & 0.002 & Yes         \\
M     & 0.029  & 0.001 & No \\
\bottomrule
\end{tabular}
\end{table}

\begin{table}\centering
\caption{Experienced segregation level ($ICE_e$) by group outside residential area. Errors are computed as the median values' bootstrap standard deviation (1000 repetitions). $^* p<0.001$ The test compares the group $ICE_e$ distributions with the random mixing thresholds defined in Section Identifying Segregated Individuals. If a group distribution is not greater than 0.2 and not smaller than -0.2, we consider it insignificant segregation (Segregation = No); otherwise, we have Segregation = Yes.}\label{tab:ice_e_stats}
\begin{tabular}{llrrl}
Group              & Scenario                                  & Median & Error  & Segregation$^*$ \\\midrule
\multirow{3}{*}{F} & Empirical                                 & -0.295 & 0.001  & Yes         \\
                   & No dest. Preference                       & -0.234 & 0.001  & Yes         \\
                   & Equalized mobility \& no dest. preference & -0.066 & 0.0003 & No          \\\midrule
\multirow{3}{*}{N} & Empirical                                 & -0.054 & 0.001  & No          \\
                   & No dest. Preference                       & -0.060 & 0.001  & No          \\
                   & Equalized mobility \& no dest. preference & -0.049 & 0.0003 & No          \\\midrule
\multirow{3}{*}{M} & Empirical                                 & -0.146 & 0.001  & No          \\
                   & No dest. Preference                       & -0.125 & 0.001  & No          \\
                   & Equalized mobility \& no dest. preference & -0.062 & 0.0002 & No \\
\bottomrule
\end{tabular}
\end{table}

\begin{table}\centering
\caption{Group exposure outside residential area ($X \rightarrow X$), share by group (\%). Errors are computed as the median values' bootstrap standard deviation (1000 repetitions). $^* $ Deviation from random mixing, $p<0.001$. The test compares the group exposure distributions with the random mixing thresholds defined in Section Identifying Segregated Individuals. For $X \rightarrow D$ and $X \rightarrow M$, if a group distribution is not greater than 19\% and not smaller than -19\%, we consider it an insignificant deviation from homogeneous exposure (Deviation = No); otherwise, we have Deviation = Yes. For $X \rightarrow F$, the criterion is not greater than 13\% and not smaller than -13\% for being insignificant.}\label{tab:inter_stats}
\begin{tabular}{llrrl}
Exposure         & Scenario                                    & Median & Error & Deviation$^*$ \\\midrule
\multirow{3}{*}{$F \rightarrow F$} & Empirical                                   & 30.0   & 0.17  & Yes                          \\
                    & No dest. Preference                         & 6.9    & 0.07  & No                           \\
                    & Equalized mobility \& no dest. preference & 4.3    & 0.01  & No                           \\\midrule
\multirow{3}{*}{$F \rightarrow N$} & Empirical                                   & -23.3  & 0.08  & Yes                          \\
                    & No dest. Preference                         & -14.3  & 0.08  & No                           \\
                    & Equalized mobility \& no dest. preference & -7.3   & 0.01  & No                           \\\midrule
\multirow{3}{*}{$N \rightarrow N$} & Empirical                                   & 17.7   & 0.14  & No                           \\
                    & No dest. Preference                         & 0.6    & 0.10  & No                           \\
                    & Equalized mobility \& no dest. preference & -1.8   & 0.02  & No                           \\\midrule
\multirow{3}{*}{$N \rightarrow F$} & Empirical                                   & -8.7   & 0.05  & No                           \\
                    & No dest. Preference                         & -3.1   & 0.02  & No                           \\
                    & Equalized mobility \& no dest. preference & 0.6    & 0.01  & No\\
\bottomrule
\end{tabular}
\end{table}

\end{document}